\documentclass[twocolumn,showpacs,preprintnumbers,
amsmath,amssymb,aps,prd,nofootinbib,superscriptaddress,
eqsecnum]{revtex4}
\usepackage{graphicx,color}

\makeatletter
\renewcommand{\p@subsection}{}
\makeatother

\renewcommand{\thesection}{\arabic{section}}

\renewcommand{\theequation}{\arabic{section}.\arabic{equation}}

\newcommand{\Slash}[1]{\ooalign%
{\hfil\rotatebox{30}{\underline{\hspace*{0.6cm}}}\hfil\crcr$#1$}}

\begin{document}

\preprint{\vbox{\hbox{DPNU-06-06}\hbox{August, 2006}}}

\title{
Thermal Dilepton Production
from Dropping $\rho$
based on the Vector Manifestation
}

\author{Masayasu Harada}
\affiliation{%
Department of Physics, Nagoya University,
Nagoya, 464-8602, Japan}
\author{Chihiro Sasaki}
\affiliation{%
Gesellschaft f\"{u}r Schwerionenforschung (GSI),
64291 Darmstadt, Germany
}

\date{\today}

\begin{abstract}
We study the pion electromagnetic form factor and the dilepton
production rate in hot matter using the hidden local 
symmetry theory 
as an effective field theory 
for pions and $\rho$ mesons.
In this framework, the chiral symmetry restoration is realized
as the vector manifestation (VM) in which the massless vector
meson becomes the chiral partner of the pion,
giving a theoretical support to the dropping $\rho$ \^^ a la
Brown-Rho scaling.
In the VM 
the vector dominance (VD) is strongly violated
near the phase transition point associated with the dropping $\rho$.
We show that
the effect of the violation of the VD
substantially suppresses the dilepton production rate
compared with the one predicted by assuming the VD together 
with the dropping $\rho$.
\end{abstract}


\setcounter{footnote}{0}

\maketitle


\section{Introduction}
\label{sec:int}

Changes of hadron properties are indications of chiral symmetry 
restoration occurring in hot and/or dense QCD and have been 
explored using various effective chiral approaches~\cite{rest,RW}.
Dileptons are considered to be promising probes since
they pass through the fireball created in heavy-ion collisions
without further interactions
and must carry information on the modifications of hadrons
in matter.
The short-lived vector mesons like the $\rho$ mesons are
expected to decay into dileptons inside the hot/dense matter.
The dilepton production around the vector meson resonances 
mainly come from two-pion processes.
An enhancement of dielectron mass spectra below
the $\rho / \omega$ resonance was first observed at CERN 
SPS~\cite{ceres} and it is an indication of the medium 
modification of the vector mesons.

The vector meson mass in matter still remains an open issue.
Although there are several scenarios
like collisional broadening due to interactions with the
surrounding hot/dense medium~\cite{RW}, 
and dropping $\rho$ meson mass associated with chiral symmetry 
restoration~\cite{BR-scaling,HY:VM},
no conclusive distinction between them has been done.
Dropping masses of hadrons following the Brown-Rho (BR)
scaling~\cite{BR-scaling} can be one of the most prominent 
candidates of the strong signal of melting quark condensate 
$\langle\bar{q}q\rangle$ which is an order parameter of 
spontaneous chiral symmetry breaking.

The vector manifestation (VM)~\cite{HY:VM} 
is a novel pattern 
of the Wigner realization of chiral symmetry in which
the $\rho$ meson becomes massless degenerate with the
pion at the chiral phase transition point.
The VM is formulated~\cite{HY:PRep,HS:VM,HKR:VM} 
in the effective field theory (EFT) based
on the hidden local symmetry (HLS)~\cite{BKUYY,BKY:PRep}.
The VM gives
a theoretical description of the dropping $\rho$ mass,
which is protected by the existence of the
fixed point (VM fixed point).

The dropping mass is supported by
the mass shift of the $\omega$ meson in nuclei measured by
the KEK-PS E325 Experiment~\cite{KEK-PS} and
the CBELSA/TAPS Collaboration~\cite{trnka}
and also that of
the $\rho$ meson observed in the STAR experiment~\cite{SB:STAR}.
Recently NA60 Collaboration provides data for the dimuon 
spectrum~\cite{NA60} which seems difficult to be explained
by a naive dropping $\rho$~\cite{NA60:2}.
However, there are still several ambiguities which are not
considered~\cite{Brown:2005ka-kb,HR,SG}.
Especially, the strong violation of the vector dominance (VD),
which is one of the significant predictions of the VM~\cite{HS:VD},
plays an important role~\cite{Brown:2005ka-kb} 
to explain the data.

In this paper,
we shall focus on the dilepton production
from the dropping $\rho$ based on the VM
using the HLS theory at finite temperature.
In the formulation of the VM,
an essential role is played by 
the {\it intrinsic temperature/density effects} of the parameters
which are introduced through the matching to QCD in the Wilsonian
sense combined with the renormalization group equations (RGEs).
We first study how the 
{\it intrinsic temperature/density effects}
affect to the dilepton spectra by comparing the predictions
of the VM with the ones obtained without intrinsic effects.
We next pay a special attention to 
the effect of the violation of the vector dominance
(indicated by ``$\Slash{\rm VD}$'') 
which is also due to the intrinsic effects.
We make a comparison of the dilepton production rates
predicted by the VM with the ones by the dropping $\rho$
with the assumption of the vector dominance.
Our result shows that the effect of the $\Slash{\rm VD}$
substantially suppresses the dilepton production rate
compared with the one predicted by assuming the VD together 
with the dropping $\rho$.

This paper is organized as follows:

In section~\ref{sec:HLS} we give a brief review on
the HLS theory and basics of the VM in matter.
In section~\ref{sec:Tdep} we determine the temperature dependences
of the parameters through the matching and RGEs.
Using those in-medium parameters, the form factor and dilepton
production rate are studied in section~\ref{sec:FF}.
There a comparison of the predictions of the VM
with those assuming the VD
together with the dropping $\rho$ is also made.
A brief summary and discussions are given in section~\ref{sec:sum}.


\setcounter{equation}{0}
\section{HLS theory and Vector Manifestation}
\label{sec:HLS} 

In this section, we briefly review the hidden local symmetry 
(HLS)~\cite{BKUYY,BKY:PRep} and the notion of the vector
manifestation (VM) of chiral symmetry
following Ref.~\cite{HY:PRep,Sasaki:D}.


\subsection{Hidden local symmetry}
\label{ssec:HLS}

The HLS Lagrangian is based on 
the $G_{\rm{global}} \times H_{\rm{local}}$ symmetry,
where $G=SU(N_f)_L \times SU(N_f)_R$ is the chiral symmetry
and $H=SU(N_f)_V$ is the HLS. 
The basic quantities are 
the HLS gauge boson and two matrix valued
variables $\xi_L(x)$ and $\xi_R(x)$
which transform as
\begin{equation}
\xi_{L,R}(x) \to \xi^{\prime}_{L,R}(x)
=h(x)\xi_{L,R}(x)g^{\dagger}_{L,R}\ ,
\end{equation}
where $h(x)\in H_{\rm{local}}\ \mbox{and}\ g_{L,R}\in
[\mbox{SU}(N_f)_{\rm L,R}]_{\rm{global}}$.
These variables are parameterized as
\begin{equation}
\xi_{L,R}(x)=e^{i\sigma (x)/{F_\sigma}}e^{\mp i\pi (x)/{F_\pi}}\ ,
\end{equation}
where $\pi = \pi^a T_a$ denotes the pseudoscalar 
Nambu-Goldstone (NG)
bosons associated with the spontaneous symmetry breaking of
$G_{\rm{global}}$ chiral symmetry, 
and $\sigma = \sigma^a T_a$ denotes
the NG bosons associated with 
the spontaneous breaking of $H_{\rm{local}}$.
This $\sigma$ is absorbed into the HLS gauge boson through 
the Higgs mechanism and the gauge boson acquires its mass.
$F_\pi$ and $F_\sigma$ are the decay constants
of the associated particles.
The phenomenologically important parameter $a$ is defined as 
\begin{equation}
a = \frac{{F_\sigma}^2}{{F_\pi}^2}\ .
\end{equation}
The covariant derivatives of $\xi_{L,R}$ are given by
\begin{eqnarray}
D_\mu \xi_L 
&=& \partial_\mu\xi_L - iV_\mu \xi_L + i\xi_L{\cal{L}}_\mu\,,
\nonumber\\
D_\mu \xi_R 
&=& \partial_\mu\xi_R - iV_\mu \xi_R + i\xi_R{\cal{R}}_\mu\,,
\end{eqnarray}
where $V_\mu$ is the gauge field of $H_{\rm{local}}$, and
${\cal{L}}_\mu \ \mbox{and}\ {\cal{R}}_\mu$ are the external
gauge fields introduced by gauging $G_{\rm{global}}$ symmetry.

In the framework of the HLS,
it is possible to perform the systematic derivative
expansion including the vector mesons as the HLS gauge boson
in addition to the pseudoscalar mesons 
as the NG bosons~\cite{Georgi,HY:PLB,Tanabashi,HY:PRep}.
In this chiral perturbation theory (ChPT) with the HLS,
the Lagrangian with lowest derivative terms 
is counted as ${\mathcal O}(p^2)$, which in the chiral limit
is given by~\cite{BKUYY,BKY:PRep}
\begin{equation}
{\cal{L}}_{(2)} 
= {F_\pi}^2\mbox{tr}\bigl[ \hat{\alpha}_{\perp\mu}
  \hat{\alpha}_{\perp}^{\mu} \bigr] 
{}+ {F_\sigma}^2\mbox{tr}\bigl[ \hat{\alpha}_{\parallel\mu}
  \hat{\alpha}_{\parallel}^{\mu} \bigr] 
{}- \frac{1}{2g^2}\mbox{tr}\bigl[ V_{\mu\nu}V^{\mu\nu} \bigr]\,,
\label{eq:L(2)}
\end{equation}
where $g$ is the HLS gauge coupling,
$V_{\mu\nu}$ is the field strength of $V_\mu$ and
\begin{eqnarray}
\hat{\alpha}_{\perp }^{\mu}
&=& \frac{1}{2i}\bigl[ D^\mu\xi_R \cdot \xi_R^{\dagger} 
{}- D^\mu\xi_L \cdot \xi_L^{\dagger} \bigr]\,,
\nonumber\\
\hat{\alpha}_{\parallel}^{\mu}
&=& \frac{1}{2i}\bigl[ D^\mu\xi_R \cdot \xi_R^{\dagger}
{}+ D^\mu\xi_L \cdot \xi_L^{\dagger} \bigr]\,.
\end{eqnarray}
There are 35 terms of ${\mathcal O}(p^4)$~\cite{Tanabashi,HY:PRep}
for general number of flavors $N_f$
(32 terms for $N_f=3$ and 24 for $N_f=2$).
Among them,
the following ${\mathcal O}(p^4)$ terms are relevant to 
the present analysis:
\begin{equation}
{\cal{L}}_{(4)} 
= z_1\mbox{tr}\bigl[ \hat{\cal{V}}_{\mu\nu}
  \hat{\cal{V}}^{\mu\nu} \bigr] 
{}+ z_2\mbox{tr}\bigl[ \hat{\cal{A}}_{\mu\nu}
  \hat{\cal{A}}^{\mu\nu} \bigr] 
{}+ z_3\mbox{tr}\bigl[ \hat{\cal{V}}_{\mu\nu}
  V^{\mu\nu} \bigr]\,, 
\label{eq:L(4)}
\end{equation}
where
\begin{eqnarray}
\hat{\cal{A}}_{\mu\nu}
=\frac{1}{2} \bigl[ \xi_R{\cal{R}}_{\mu\nu}\xi_R^{\dagger}
{}- \xi_L{\cal{L}}_{\mu\nu}\xi_L^{\dagger} \bigr]\,,
\label{def A mn}
\nonumber\\
\hat{\cal{V}}_{\mu\nu}
=\frac{1}{2} \bigl[ \xi_R{\cal{R}}_{\mu\nu}\xi_R^{\dagger}
{}+ \xi_L{\cal{L}}_{\mu\nu}\xi_L^{\dagger} \bigr]\,,
\label{def V mn}
\end{eqnarray}
with ${\cal{R}}_{\mu\nu}$ and ${\cal{L}}_{\mu\nu}$ being
the field strengths of external gauge fields
${\cal{R}}_{\mu}$ and ${\cal{L}}_{\mu}$.


\subsection{Wilsonian matching}
\label{ssec:match}

The Wilsonian matching was proposed as a novel manner
to determine the parameters of
effective field theories (EFTs) 
from the underlying QCD~\cite{HY:WM}.
It was applied to the HLS theory in the vacuum, 
which gives several predictions
in good agreement with experiments~\cite{HY:WM,HY:PRep}.
The Wilsonian matching 
has been applied to study 
chiral phase transition at a large number of 
flavor~\cite{HY:VM,HY:PRep} and at finite 
temperature/density~\cite{HS:VM,HKR:VM}.
The matching in the Wilsonian sense is performed based on the 
following general idea:
The bare Lagrangian of an EFT is defined at a suitable 
high energy scale $\Lambda$ and
the generating functional derived from the bare Lagrangian 
leads to the same Green's function as that derived from original
QCD Lagrangian at $\Lambda$.
Then the {\it bare} parameters of the EFT are determined 
through the Wilsonian matching.
In other words,
one obtains the bare Lagrangian of the EFT after
integrating out the high energy modes, i.e.,
the quarks and gluons above $\Lambda$.
The information of the high energy modes is included in the 
parameters of the EFT.

We first show the basics of the Wilsonian matching in the vacuum.
The Wilsonian matching is done by matching
the axial-vector and vector current correlators derived from the
HLS with those by the operator product expansion (OPE) in
QCD at the matching scale $\Lambda$.
For the validity of the expansion in the HLS the
matching scale $\Lambda$ must be smaller than the chiral symmetry
breaking scale $\Lambda_\chi$, above which the ChPT with HLS breaks
down.
On the other hand, the matching scale
$\Lambda$ should be small enough for the
validity of the OPE.
In Refs.~\cite{HY:WM,HY:PRep,FHS}, several choices of $\Lambda$ around
$1.1\,\mbox{GeV}$ were shown to provide good predictions on 
the low-energy 
phenomenology related with the $\rho$ meson in real-life QCD.

In the OPE,
the axial-vector and vector correlators are expressed
as~\cite{SVZ}
\begin{eqnarray}
&&
G^{\rm{(QCD)}}_A (Q^2) 
= \frac{1}{8{\pi}^2}\Bigl[ -
    \bigl( 1 + \frac{\alpha _s}{\pi} \bigr)
    \ln \frac{Q^2}{{\mu}^2} 
\nonumber\\
&&\qquad
{}+ \frac{{\pi}^2}{3}
    \frac{\langle \frac{\alpha _s}{\pi}
    G_{\mu \nu}G^{\mu \nu} \rangle }{Q^4} 
{}+
    \frac{{\pi}^3}{3} \frac{1408}{27}
    \frac{\alpha _s{\langle \bar{q}q
    \rangle }^2}{Q^6} \Bigr]\,, 
\nonumber\\
&&
G^{\rm{(QCD)}}_V (Q^2) 
= \frac{1}{8{\pi}^2}\Bigl[ -
    \bigl( 1 + \frac{\alpha _s}{\pi} \bigr)
    \ln \frac{Q^2}{{\mu}^2}
\nonumber\\
&&\qquad
{}+ \frac{{\pi}^2}{3}
    \frac{\langle \frac{\alpha _s}{\pi}
    G_{\mu \nu}G^{\mu \nu} \rangle }{Q^4} -
    \frac{{\pi}^3}{3} \frac{896}{27}
    \frac{\alpha _s{\langle \bar{q}q
    \rangle }^2}{Q^6} \Bigr]\,,
\label{CC-OPE-vac}
\end{eqnarray}
where $\mu$ is the renormalization scale.
These current correlators in the HLS around 
the matching scale $\Lambda$ are well described by the following 
forms with the bare parameters:
\begin{eqnarray}
&&
G^{\rm{(HLS)}}_A (Q^2) 
= \frac{F^2_\pi (\Lambda)}{Q^2} -
    2z_2(\Lambda)\,, 
\nonumber\\
&&
G^{\rm{(HLS)}}_V (Q^2) 
\nonumber\\
&&\qquad
= \frac{F^2_\sigma (\Lambda)[1 - 2g^2(\Lambda)z_3(\Lambda)]}
    {{M_\rho}^2(\Lambda) + Q^2} - 2z_1(\Lambda)\,.
\label{CC-HLS-vac}
\end{eqnarray}
We set the above correlators to be equal to those 
in Eq.~(\ref{CC-OPE-vac}) at $\Lambda$
up until the first derivative to obtain the Wilsonian matching 
conditions at zero temperature~\cite{HY:WM}.
Quantum corrections are incorporated into the parameters
through the renormalization group equations (RGEs).
Within this framework, several physical quantities
in the low-energy region have been studied and the predictions
of the Wilsonian matching are in good agreement with experiments.
For details, see Refs.~\cite{HY:WM,HY:PRep}.

\begin{figure}
\begin{center}
\includegraphics[width = 7cm]{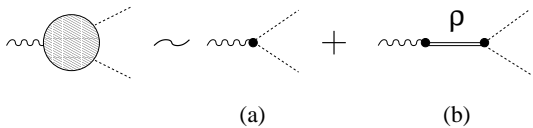}
\end{center}
\caption{Leading contributions to the electromagnetic form factor
 of the pion; (a) the direct $\gamma\pi\pi$ interaction and 
 (b) the $\gamma\pi\pi$ interaction mediated by 
 the $\rho$-meson exchange.
}
\label{fig:VD}
\end{figure}

For showing a typical prediction of the Wilsonian matching
in the vacuum, 
let us consider the pion electromagnetic form factor.
The leading contributions to the pion form factor are
shown in Fig.~\ref{fig:VD}.
The form factor is expressed as
\begin{eqnarray}
&&
{\cal F}(s)
= g_{\gamma\pi\pi} 
\nonumber\\
&&\quad
{}+ \frac{g_\rho(s) \cdot g_{\rho\pi\pi}}
    {m_\rho^2 - s - \theta(s - 4m_\pi^2) 
     i m_\rho \Gamma_\rho(s)}\,,
\label{form factor-vac}
\end{eqnarray}
where $\theta$ denotes the step function.
In the above
$g_{\gamma\pi\pi}$ is the contribution
from the direct photon-$\pi$-$\pi$ interaction, and the remaining
is the one mediated by the $\rho$ meson exchange.
The direct photon-$\pi$-$\pi$ coupling constant $g_{\gamma\pi\pi}$
in the HLS is given as
\begin{equation}
g_{\gamma\pi\pi} = 1 - \frac{a(0)}{2} \,,
\label{gamma-pi-pi}
\end{equation}
with $a(0)$ being the parameter evaluated at $\mu=0$.
$g_{\rho\pi\pi}$ is the $\rho\pi\pi$ coupling given by
\begin{equation}
g_{\rho\pi\pi} = g(m_\rho)\,\frac{a(0)}{2}
\label{grpp_vac}
\end{equation}
where $g(m_\rho)$ is the HLS gauge coupling at
the $\rho$ on-shell, $\mu = m_\rho$.
$g_\rho(s)$ is the momentum dependent $\rho$-$\gamma$ mixing strength
given by
\begin{equation}
g_\rho(s) = \frac{m_\rho^2}{g(m_\rho)} - s\, g(m_\rho) z_3(m_\rho)
\ ,
\label{gr_vac}
\end{equation}
where $z_3(m_\rho)$ is one of the ${\mathcal O}(p^4)$ parameters
given in Eq.~(\ref{eq:L(4)}).
$\Gamma_\rho(s)$ is the momentum dependent $\rho$ width 
given by
\begin{eqnarray}
\Gamma_\rho(s)
&=& \frac{m_\rho}{\sqrt{s}}
\left( \frac{s - 4m_\pi^2}{m_\rho^2 - 4m_\pi^2} \right)^{3/2}
\Gamma_\rho\,,
\nonumber\\
\Gamma_\rho
&=& \frac{1}{6\pi m_\rho^2} 
\left( \frac{m_\rho^2 - 4m_\pi^2}{4} \right)^{3/2}
\left| g_{\rho\pi\pi} \right|^2\,.
\nonumber\\
\end{eqnarray}

\begin{figure}
\begin{center}
\includegraphics[width = 7cm]{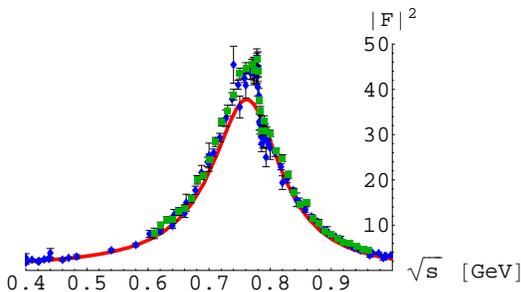}
\end{center}
\caption{
Electromagnetic form factor of the pion as a function of
the invariant mass $\sqrt{s}$ at zero temperature.
The experimental data were taken from 
Ref.~\cite{exp1} (indicated by $\diamond$)
and Ref.~\cite{exp2} (by $\square$).
}
\label{fig:form-vac}
\end{figure}

We perform the Wilsonian matching as shown
in Ref.~\cite{HY:PRep} using 
the $\rho$ mass, $m_\rho = 775.8\,\mbox{MeV}$, and
the pion decay constant, $f_\pi = 92.46\,\mbox{MeV}$, as inputs
and determine the parameters of the HLS as~\footnote{
  In Ref.~\cite{HY:PRep}, $m_\rho = 771.1\,\mbox{MeV}$ and
  the pion decay constant at the chiral limit,
  $F_\pi(0) = 86.4\pm 9.7\,\mbox{MeV}$, were used as inputs.
}
\begin{eqnarray}
a(0) &=& 1.94 \,, \nonumber\\
g(m_\rho) &=& 6.02 \,, \nonumber\\
z_3(m_\rho) &=& -3.87 \times 10^{-3} \,,
\label{wm_vac}
\end{eqnarray}
for 
$\Lambda_{\rm QCD}=0.4\,\mbox{GeV}$ and 
$\Lambda=1.1\,\mbox{GeV}$.
In Fig.~\ref{fig:form-vac} we 
show the pion form factor in the time-like region
predicted from the Wilsonian matching
together with the experimental data obtained from
the $e^+e^- \to \pi\pi$ process.
We note that an extra peak of the data 
at $\sqrt{s} \sim 0.78$\,GeV is caused by the mixing of 
the $\rho$ meson to the $\omega$ meson
due to the small isospin symmetry breaking,
which is neglected in the present analysis.
It can be seen that the Wilsonian matching well describes
the experiment except for the $\omega$ region.

Next we apply this procedure to the study of hot/dense matter.
As we noted in the beginning of this subsection,
the bare parameters are determined by integrating out high
frequency modes above the matching scale.
Thus when we integrate out those degrees of freedom 
in hot/dense matter,
the bare parameters are dependent on temperature/density.
We shall refer them as the 
{\it intrinsic temperature/density effects}~\cite{HS:VM,HKR:VM}.
The intrinsic temperature and/or density dependences are 
nothing but the signature that the hadron has an internal structure 
constructed from the quarks and gluons.
This is similar to the situation where the coupling constants
among hadrons are replaced with the momentum-dependent form factor
in the high-energy region.
Thus the intrinsic temperature/density effects 
play more important roles in higher temperature/density 
region, especially near the phase transition point.

It should be noticed that
there is no longer Lorentz symmetry in matter,
and the Lorentz non-scalar operators such as
$\bar{q}\gamma_\mu D_\nu q$ may exist in 
the expressions of the OPE current correlators.
However, we neglect the contributions from these operators
since such effects are suppressed by powers of the matching 
scale~\cite{HKRS:SUS,PiV}.
Thus it is a good approximation that we determine the bare parameters
at non-zero temperature/density through the matching 
conditions with putting possible 
temperature/density dependences on the quark and gluon 
condensates:
\begin{eqnarray}
&
\langle \bar{q}q \rangle \to
\langle \bar{q}q \rangle_{T,\mu_q}\,,
&
\nonumber\\
&
\langle G_{\mu\nu}G^{\mu\nu} \rangle \to
\langle G_{\mu\nu}G^{\mu\nu} \rangle_{T,\mu_q}\,.
&
\label{ope-matter}
\end{eqnarray}
Through the matching conditions,
the temperature/density dependences of those condensates
determine the intrinsic effects of the bare parameters,
which are then converted into those of the on-shell parameters
through the Wilsonian RGEs.


\subsection{VM in hot/dense matter}
\label{ssec:VM} 

The vector manifestation (VM) was proposed in Ref.~\cite{HY:VM}
as a novel pattern of the Wigner realization of chiral symmetry
with a large number of massless quark flavors,
in which the vector meson becomes massless at the restoration
point and belongs to the same chiral multiplet as the pion,
i.e., {\it the massless vector meson is the chiral partner 
of the pion}.
The studies of the VM in hot/dense matter have been carried out
in Refs.~\cite{HS:VM,HKR:VM,HS:VD,HKRS:SUS,PiV} and the VM was
applied to construct an effective Lagrangian for the heavy-light 
mesons which can describe the recent experimental observation 
on the $D(0^+,1^+)$ mesons~\cite{HRS:CD}.
We would like to stress that
the vector dominance (VD) of the electromagnetic
form factor of the pion is strongly violated near
the critical point associated with the dropping $\rho$
in the VM~\cite{HY:VD,HS:VD}.
In the following, we briefly review the VM in hot/dense matter.

We consider the quantity $G_A - G_V$ which is a measure of the 
spontaneous chiral symmetry breaking in QCD.
Here we assume that the quark condensate approaches zero
when the restoration point is approached from the broken phase:
The phase transition is of second order.
In such a case, as we can easily see in Eq.~(\ref{CC-OPE-vac}),
$G_A - G_V$ in the OPE approaches zero as 
\begin{eqnarray}
&&
G^{\rm{(QCD)}}_A (Q^2) - G^{\rm{(QCD)}}_V (Q^2) 
\nonumber\\
&& \quad
=
\frac{32\pi}{9} \frac{\alpha_s \langle\bar{q}q\rangle^2}{\Lambda^6}
\ \rightarrow \ 0 \,.
\label{cc OPE diff}
\end{eqnarray}
This condition $G_A - G_V \to 0$
should be also satisfied in the EFT.
We thus impose that these current correlators approach each 
other at the bare level, which implies
\begin{equation}
G_A^{\rm (HLS)}(\Lambda;T,\mu_q) - G_V^{\rm (HLS)}(\Lambda;T,\mu_q) 
\to 0\,,
\end{equation}
for $(T,\mu_q) \to (T_c,\mu_q^c)$.
We require that this equality is satisfied at an arbitrary momentum
scale $Q$ near the matching scale $\Lambda$.
This can be satisfied only when the following conditions for the bare
parameters are met:
\begin{eqnarray}
&
g(\Lambda;T,\mu_q) \to 0\,, \quad a(\Lambda;T,\mu_q) \to 1\,
&
\nonumber\\
&
z_1(\Lambda;T,\mu_q) - z_2(\Lambda;T,\mu_q) \to 0\,,
&
\label{vm-cond}
\end{eqnarray}
for $(T,\mu_q) \to (T_c,\mu_q^c)$.
The first condition immediately implies that the bare $\rho$ mass
approaches zero: $M_\rho(\Lambda;T,\mu_q) \to 0$.
The second one leads to the agreement of two decay constants:
$F_\sigma(\Lambda;T,\mu_q) - F_\pi(\Lambda;T,\mu_q) \to 0$.

We should note that Eq~(\ref{vm-cond}) is realized due to 
the intrinsic $T/\mu_q$ effects introduced
through the Wilsonian matching.
It was shown~\cite{HY:PRep,HS:VM,HKR:VM}
that these conditions are protected by the fixed point of 
the RGEs (see also Eq.~(\ref{rge}) in the next section)
and never receives 
quantum corrections at the critical point.
Thus the parametric vector meson mass determined at 
the on-shell of the vector meson also vanishes since it is
proportional to the vanishing gauge coupling constant.
The vector meson mass $m_\rho$ defined as a pole position of 
the full vector meson propagator has the hadronic corrections 
through thermal/dense loops, which are proportional to
the gauge coupling constant~\cite{HS:VM,HKR:VM,HS:VD}.
Consequently the vector meson pole mass also goes to zero
for $(T,\mu_q) \to (T_c,\mu_q^c)$:
\begin{equation}
m_\rho(T,\mu_q) \to 0\,.
\end{equation}
By performing the matching to $G_A^{\rm (OPE)}-G_V^{\rm (OPE)}$ 
near the critical point and using the fact that the vanishing 
gauge coupling constant is the fixed point of the RGE, 
we find that {\it the $\rho$ meson mass
drops as the chiral condensate};
\begin{equation}
m_\rho(T,\mu_q) \propto \langle \bar{q}q \rangle_{T,\mu_q}\,,
\end{equation}
for $(T,\mu_q) \simeq (T_c,\mu_q^c)$.
Note that the Wilsonian matching condition at the critical point
provides the non-zero {\it bare} pion decay constant,
$F^2_\pi (\Lambda;T_c,\mu_q^c) \neq 0$, 
even at the critical point where the on-shell pion decay constant 
vanishes by adding the quantum corrections through
the RGEs including the quadratic divergences~\cite{HY:VM}
and hadronic temperature/density corrections~\cite{HS:VM,HKR:VM}.

It has been shown that the vector dominance (VD) 
of the electromagnetic form factor of the pion~\cite{Sakurai}
is accidentally satisfied in $N_f=3$ QCD 
at zero temperature and zero density, and that it is strongly 
violated in large $N_f$ QCD when the VM occurs~\cite{HY:VD}.
The VD is characterized by the direct $\gamma\pi\pi$ being zero.
As one can easily see from Eq.~(\ref{gamma-pi-pi}),
the VD can be well satisfied when the parameter $a$ is close to
$2$; $a \simeq 2$.
This is actually predicted by the Wilsonian matching 
in the vacuum as presented in Eq.~(\ref{wm_vac}).
In hot/dense matter, the parameter $a$ is modified
by medium effects and approaches unity with increasing $T/\mu_q$ 
toward the critical point, which is due to the intrinsic
$T/\mu_q$ effects associated with the chiral symmetry restoration
~\cite{HKR:VM,HS:VD}.
This implies that {\it the VD is strongly violated near the critical 
point, maximally 50 \%}, and it
strongly affects the understanding of experiments
on the dilepton productions based on the dropping $\rho$
as recently pointed out in Ref.~\cite{Brown:2005ka-kb}.
In the following sections, we will carry out an analysis of 
the spectral function and the dilepton production rate taking into
account of the strong violation of the VD.


\setcounter{equation}{0}
\section{Temperature dependence of parameters}
\label{sec:Tdep} 

Generally, the parameters of effective field theories
(EFTs) in hot/dense matter have dependences on 
the temperature/density.
This is nothing but the signature that the particles 
expressed by effective fields are composites of the fundamental
degrees of freedom.
In the HLS, such dependences (intrinsic
temperature/density dependences) can be introduced at the bare level 
through the Wilsonian matching.
The physical quantities in the low-energy region
then have two kinds of temperature/density dependences, 
one is the intrinsic effect and another
comes from the ordinary hadronic corrections included through
the temperature/density loops.
In the following analysis to determine both effects, 
we consider the system at finite temperature and zero density.

We consider the intrinsic temperature dependence
of the bare parameters of the HLS Lagrangian based on the
Wilsonian matching.
With increasing temperature toward
the critical temperature, the difference of two
current correlators in the OPE approaches zero 
as in Eq.~(\ref{cc OPE diff}).
Noting that
the Wilsonian matching condition 
obtained from the first derivative of the axial-vector 
current correlator provides
\begin{eqnarray}
\frac{F_\pi^2(\Lambda;T_c)}{\Lambda^2}
=
\frac{1}{8\pi^2}
\Biggl[
  1 + \frac{\alpha_s}{\pi}
  \frac{2\pi^2}{3} 
  \frac{ 
    \left\langle
      \frac{\alpha_s}{\pi} G_{\mu\nu}G^{\mu\nu} 
    \right\rangle_{T_c}
  }{ \Lambda^4 }
\Biggr]
\neq 0 \,,
\nonumber\\
\label{WM:Fpi:Tc}
\end{eqnarray}
even at the critical point,
we can expand the difference of two correlators
$G_A - G_V$ as
\begin{eqnarray}
&&
G_A^{\rm(HLS)}(Q^2) - G_V^{\rm(HLS)}(Q^2)
\nonumber\\
&&\quad \simeq
  g^2(\Lambda;T)\, 
   \left( \frac{F_\pi^2(\Lambda;T)}{\Lambda^2} \right)^2
\nonumber\\
&& \qquad
  - \left( a(\Lambda) - 1 \right)\,
   \frac{F_\pi^2(\Lambda;T)}{\Lambda^2}
\nonumber\\
&& \qquad
 + 2 g^2(\Lambda;T) \, z_3(\Lambda;T) \,
   \frac{F_\pi^2(\Lambda;T)}{\Lambda^2}
\nonumber\\
&& \qquad
  - 2 \left( z_2(\Lambda;T) - z_1(\Lambda;T) \right)
\ ,
\label{cc HLS diff}
\end{eqnarray}
near the critical point based on the VM.
Comparing Eq.~(\ref{cc HLS diff}) with Eq.~(\ref{cc OPE diff})
and requiring no cancellations among the terms in the right-hand-side
of Eq.~(\ref{cc HLS diff}),
we see that
both bare $g$ and 
$a-1$ are proportional to the quark condensate
provided by the Wilsonian matching near $T_c$~\cite{HS:VM}:
\begin{eqnarray}
&&
g^2(\Lambda;T) \propto \langle \bar{q}q \rangle_{T}^2
\nonumber\\
&&
a(\Lambda;T) - 1 \propto \langle \bar{q}q \rangle_{T}^2
\quad \mbox{for}\,\, T \simeq T_c\,.
\label{match-input}
\end{eqnarray}
This implies that the bare parameters are thermally evolved
following the temperature dependence of the quark condensate,
which is nothing but the intrinsic temperature effect.

We should note that
the above matching conditions hold
{\it only in the vicinity of $T_c$}:
Equation~(\ref{match-input}) is not valid any more far away from
$T_c$ where ordinary hadronic corrections are dominant.
For expressing a temperature above which the intrinsic
effect becomes important,
we introduce a temperature $T_f$, 
so-called flash temperature~\cite{BLR:flash,BLR}.
The VM and therefore the dropping $\rho$ mass become 
transparent for $T>T_f$.
On the other hand, we expect that
the intrinsic effects are negligible in the low-temperature
region below $T_f$:
Only hadronic thermal corrections are considered for $T < T_f$.
Based on the above consideration, we adopt the following
ansatz of the temperature dependences of the 
bare $g$ and $a$:~\footnote{
 As was stressed in Refs.~\cite{HY:PRep,Sasaki:D}, the VM should be
 considered only as the limit.  So we include the temperature
 dependences of the parameters only for $T_f < T < T_c - \epsilon$.
}
\begin{eqnarray}
&&
\mbox{for}\,T < T_f \quad
\left\{\begin{array}{l}
  g(\Lambda;T) = \mbox{(constant)} 
 \\
  a(\Lambda;T) - 1 = \mbox{(constant)} 
\end{array} \right.
\nonumber\\
&&
\mbox{for}\,T > T_f \quad
\left\{\begin{array}{l}
  g(\Lambda;T) \propto \langle \bar{q}q \rangle_{T} 
 \\
  a(\Lambda;T) - 1 \propto \langle \bar{q}q \rangle_{T}^2
\end{array} \right.
\label{bare t dep}
\end{eqnarray}

In the following analysis,
we assume that the quark condensate scale as
$(T_c - T)^{1/2}$ as in the mean field case as an example.
Then, the above ansatz in Eq.~(\ref{bare t dep})
gives the following temperature dependences of the bare $g$ 
and $a$, which approach the VM fixed point near $T_c$:
\begin{eqnarray}
&&
g(\Lambda;T)
= 
\theta(T_f - T) g(\Lambda;T=0) 
\nonumber\\
&&\quad
{}+ \theta(T - T_f)g(\Lambda;T=0)
\left( 1 - \frac{T^2 - T_f^2}{T_c^2 - T_f^2} \right)^{1/2}\,,
\nonumber\\
&&
a(\Lambda;T)
= 
\theta(T_f - T) a(\Lambda;T=0) 
{}+ \theta(T - T_f)
\nonumber\\
&&\times
\left[ 1 + \left\{ a(\Lambda;T=0) - 1 \right\}
\left( 1 - \frac{T^2 - T_f^2}{T_c^2 - T_f^2} \right)\right]\,,
\label{Tdep g a}
\end{eqnarray}
where $\theta$ is the step function.
Although the matching scale $\Lambda$ can also depend on 
temperature in hot matter, we assume that
it does not have any temperature dependence in the
present analysis.
Since the Wilsonian matching conditions
[see Eq.~(\ref{WM:Fpi:Tc}) and Eq.~(\ref{cc HLS diff}) 
together with Eq.~(\ref{cc OPE diff})]
do not provide a substantial
temperature dependence on 
the ratio
$F_\pi(\Lambda;T)/\Lambda$
as well as 
the parameter
$z_3(\Lambda;T)$,
we assume that the bare $z_3$ and bare $F_\pi$ 
do not have any $T$ dependences:
\begin{eqnarray}
&&
z_3(\Lambda;T) = z_3(\Lambda;0)\,,
\nonumber\\
&&
F_\pi(\Lambda;T) = F_\pi(\Lambda;0)\,.
\label{Tdep z F}
\end{eqnarray}

For making a numerical analysis including hadronic
corrections in addition to the intrinsic effects determined
above, we take
$T_c = 170\,\mbox{MeV}$ as a typical example
and $T_f = 0.7\,T_c$ as proposed in Refs.~\cite{BLR:flash,BLR},
where the value of $T_f$ is fixed
based on the following consideration:
The gluon condensate which leads to the trace anomaly 
in QCD, is divided
into the soft glue as a mean field and the hard glue as
the fluctuation.
The soft glue gives spontaneous breaking of scale invariance
and the hard glue explicitly breaks it.
The anomalous breaking from quark part is proportional to 
$m_q \bar{q}q$ with $m_q$ being the current quark mass which 
indeed vanishes in the classical level if $m_q = 0$.
The soft glue melts toward the chiral phase transition point
and eventually the scale invariance is restored at $T_c$.
[Although the scale invariance is still explicitly broken by
 the hard glue, it has nothing to do with hadron masses.]
In Ref.~\cite{BR-scaling} an effective chiral Lagrangian which
reproduces the proper scaling behavior even in the level of the EFT
were presented and the BR scaling was proposed based on the notion
of the melting soft glue.
The BR scaling deals with the quantity directly locked to the 
quark condensate and hence {\it the scaling masses are achieved 
exclusively by the intrinsic effect.}
According to the lattice calculation~\cite{lattice-glue},
the gluon condensate starts to drop around $T = 0.7\,T_c$
and remains finite at $T_c$ which is about half of that at $T=0$.
The melting soft glue is due to the intrinsic temperature effect
which comes in at $T_f$.

\begin{figure}
\begin{center}
\includegraphics[width = 7cm]{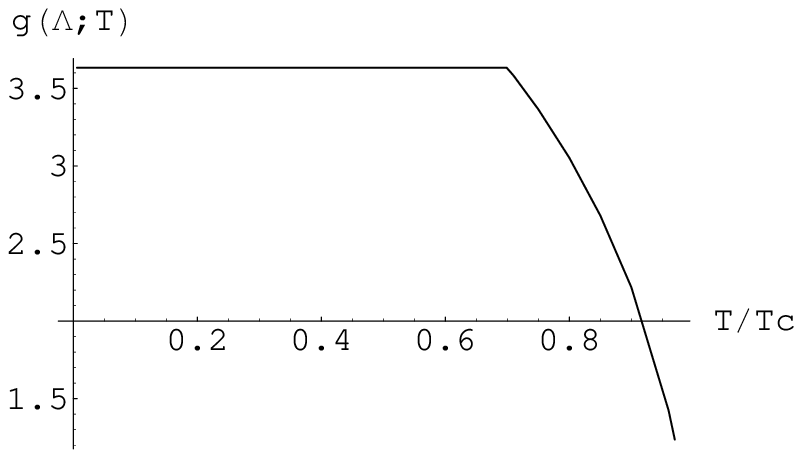}
\\
(a)
\\
\includegraphics[width = 7cm]{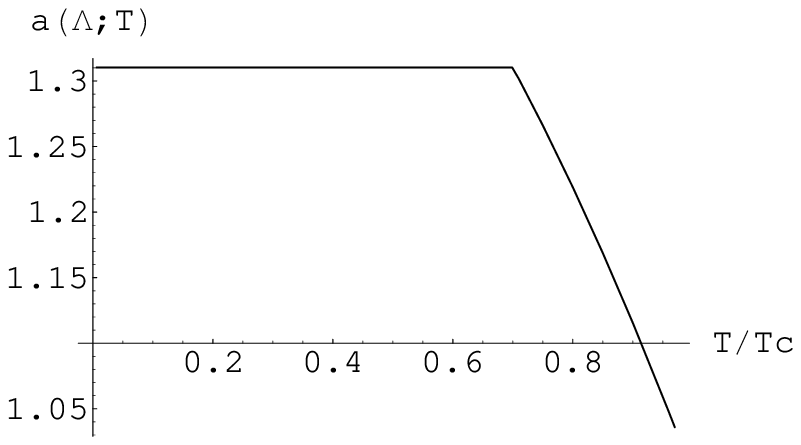}
\\
(b)
\end{center}
\caption{
Temperature dependences of (a) the bare gauge coupling constant
$g$ and (b) the bare $a$.
}
\label{fig:bareT}
\end{figure}

For the bare parameters of the HLS Lagrangian at $T=0$,
we use the ones determined through the Wilsonian matching
for 
$\Lambda_{\rm QCD}=0.4\,\mbox{GeV}$ and $\Lambda=1.1\,\mbox{GeV}$
($N_c = 3$ and $N_f = 3$),
which well describe~\cite{HY:WM,HY:PRep} the experiments 
at $T=0$ as briefly shown in subsection~\ref{ssec:match}.
For later convenience, we summarize
our input parameters in Table~\ref{table:input}.
\begin{table}
\begin{center}
\begin{tabular}{|c|c|c|c|}
\hline
$f_\pi$ [GeV] & 
$m_\rho$ [GeV] &  
$\Lambda_{\rm QCD}$ [GeV] &
$\Lambda$ [GeV]
\\
\hline
$0.09246$ & $0.7758$ & $0.4$ & $1.1$
\\
\hline\hline
$F_\pi(\Lambda)$ [GeV] &
$a(\Lambda)$ &
$g(\Lambda)$ &
$z_3(\Lambda) \times 10^3$
\\
\hline
$0.149$ & $1.31$ & $3.63$ & $-2.55$
\\
\hline\hline
$m_\pi$ [GeV] &
$T_c$ [GeV] &
$T_f$ &
{---}
\\
\hline
$0.13957$ & $0.17$ & $0.7\,T_c$ & {---}
\\
\hline
\end{tabular}
\end{center}
\caption{
 Input parameters in the present analysis.
 The values listed in the second line are
 fixed by performing the Wilsonian
 matching at $T=0$ with $\langle \bar{q}q \rangle_{\rm 1 GeV}
 = - (0.25)^3$ GeV$^3$ and $\langle \frac{\alpha_s}{\pi}G_{\mu\nu}
 G^{\mu\nu} \rangle = 0.012$ GeV$^4$.
}
\label{table:input}
\end{table}
Note that the intrinsic effects also generate the Lorentz 
non-invariance for the bare parameters and thus the on-shell 
parameters, which were neglected here since such effects are
much suppressed by the matching scale~\cite{PiV}.
We show the temperature dependences of
the bare parameters $g$ and $a$ in Fig.~\ref{fig:bareT}.

Next we incorporate quantum corrections into the parameters
following the RGEs at one loop~\cite{HY:WM}:
\begin{eqnarray}
\mu \frac{d{F_\pi}^2}{d\mu} 
&=& 
\frac{N_f}{2(4\pi)^2}
\left[ 3a^2 g^2 {F_\pi}^2 + 2(2-a){\mu}^2 \right]\,,
\nonumber\\
\mu \frac{d a}{d\mu} 
&=& 
{}- \frac{N_f}{2(4\pi)^2}(a-1)
\nonumber\\
&&\times
\left[ 3a(a+1)g^2 - (3a-1)\frac{{\mu}^2}{{F_\pi}^2} \right]\,,
\nonumber\\
\mu \frac{dg^2}{d\mu} 
&=& 
{}-\frac{N_f}{2(4\pi)^2}\frac{87-a^2}{6}g^4\,,
\nonumber\\
\mu \frac{d z_3}{d\mu}
&=&
\frac{N_f}{(4\pi)^2}\frac{1 + 2 a - a^2}{12}\,.
\label{rge}
\end{eqnarray}
These parameters are evolved from the matching scale to their 
on-shell and the parameters at the physical scale are obtained.
In the following we put the bar to the parameters of the
Lagrangian to clarify
that the barred quantities
include the intrinsic effects and are renormalized at on-shell
of relevant particles: e.g.,
\begin{eqnarray}
\bar{g} &=& g(\mu=m_\rho;T)\,,
\nonumber\\
\bar{a} &=& a(\mu=m_\rho;T)\,,
\nonumber\\
\bar{z}_3 &=& z_3(\mu=m_\rho;T)\,,
\nonumber\\
\bar{F}_\pi &=& F_\pi(\mu=0;T)\,,
\nonumber\\
\bar{F}_\sigma &=& \sqrt{\bar{a}} F_\pi(\mu=m_\rho;T)\,,
\nonumber\\
\bar{M}_\rho &=& \bar{g} \bar{F}_\sigma \,.
\end{eqnarray}

Physical quantities are obtained by
including the hadronic corrections generated through thermal
loop diagrams at one loop.
The hadronic correction to the vector meson mass
is given by~\cite{HS:VM,HS:VD}%
~\footnote{
 In the following analysis, we include the hadronic thermal
 corrections only for the leading order parameters, 
 $F_\pi, F_\sigma$ and $g$. For the calculation of the form factor
 ${\cal F}$, $z_2 - z_1$ is irrelevant. 
 Although $g_\rho(s)$ includes $z_3$, by which a 30 \% deviation
 of experiments from the KSRF I relation can be explained 
 in the vacuum~\cite{HY:WM,HY:PRep}, 
 we will neglect the hadronic correction to it.
}
\begin{eqnarray}
&&
m_\rho^2(T)
= \bar{M}_\rho^2 + N_f \bar{g}^2
  \Bigl[ - \frac{\bar{a}^2}{12}G_2(\bar{M}_\rho;T)
\nonumber\\
&&\quad
{}+ \frac{5}{4}J_1^2(\bar{M}_\rho;T) 
{}+ \frac{33}{16}\bar{M}_\rho^2 
F_3^2(\bar{M}_\rho;\bar{M}_\rho;T) \Bigr]\,,
\end{eqnarray}
where the functions $J, G$ and $F$ are listed 
in Appendix~\ref{app:functions},
and $m_\rho$ is defined at the
rest frame of the $\rho$ meson, 
$p_\mu = (m_\rho, \vec{0})$%
~\footnote{
 In medium the full vector meson propagator consists of
 the longitudinal and tranverse parts, $D_L$ and $D_T$.
 The vector meson masses defined as the pole positions of $D_L$
 and $D_T$ are identical at rest frame.
},

Differently from the vacuum, the Lorentz invariance is not 
manifest any more due to the heat bath.
The in-medium pion decay constants, the temporal and spatial
components $f_\pi^{t,s}$ are defined by~\cite{Pisarski}
\begin{eqnarray}
&&
\langle 0 | J_5^0 | \pi(p) \rangle
= i f_\pi^t p^0\,,
\nonumber\\
&&
\langle 0 | J_5^i | \pi(p) \rangle
= i f_\pi^s p^i\,,
\end{eqnarray}
where $J_5^\mu$ denotes the axial-vector current.
The order parameter of the chiral symmetry is defined as
the pole residue of the axial-vector current correlator
following Ref.~\cite{fpi-def},
which provides $f_\pi^t f_\pi^s$~\cite{HKRS:SUS}.
On the other hand, the $f_\pi^t$ is the wave function 
renormalization constant of the $\pi$ field~\cite{MOW,HKRS:SUS}.
The hadronic corrections to two pion decay constants 
are~\cite{HS:VM,HS:VD}
\begin{eqnarray}
&&
\left( f_\pi^t(T) \right)^2 
= \bar{F}_\pi^2 
\nonumber\\
&&\quad
{}- N_f \left[ I_2(T) - \frac{\bar{a}}{\bar{M}_\rho^2}
\left( I_4(T) - J_1^4(\bar{M}_\rho;T) \right) \right]\,,
\nonumber\\
&&
f_\pi^t(T)f_\pi^s(T) 
= \bar{F}_\pi^2 - N_f \Biggl[ I_2(T)
\nonumber\\
&&\quad 
{}+ a \left\{ \frac{1}{3\bar{M}_\rho^2}
\left( I_4(T) - J_1^4(\bar{M}_\rho;T) \right) - J_1^2(\bar{M}_\rho;T) 
\right\} \Biggr]\,,
\nonumber\\
\end{eqnarray}
where the functions $I$ and $J$ are listed 
in Appendix~\ref{app:functions} and the soft pion limit 
for $f_{\pi}^{t,s}$ was taken.

In general there are two decay constants for $\sigma$ 
(longitudinal $\rho$).
However, they become identical to each other when 
the bare Lagrangian has the Lorentz invariance:
\begin{eqnarray}
&&
f_\sigma(T)^2 =
\left( f_\sigma^t(T) \right)^2
= f_\sigma^t(T)f_\sigma^s(T)
\nonumber\\
&&\quad
= \bar{F}_\sigma^2 
  - \frac{N_f}{4}\left[ \bar{a}^2 I_2(T) - J_1^2(\bar{M}_\rho;T)
{}+ 2 J_{-1}^0(\bar{M}_\rho;T) \right]\,.
\nonumber\\
\label{fsig:T}
\end{eqnarray}
As we show in Appendix~\ref{app:Imaginary} (see Eq.~(\ref{app:grho})),
the momentum dependent $\rho$-$\gamma$ mixing strength is given by
\begin{eqnarray}
&&
g_\rho(s;T) = \bar{g}\left[ 
  f_\sigma^2(T) - s\, \bar{z}_3 \right] \,,
\label{gr}
\end{eqnarray}
where $f_\sigma^2(T)$ includes both intrinsic and 
hadronic temperature effects as in Eq.~(\ref{fsig:T}), 
while $\bar{g}$ and $\bar{z}_3$
include only the intrinsic one.

\begin{figure}
\begin{center}
\includegraphics[width = 7cm]{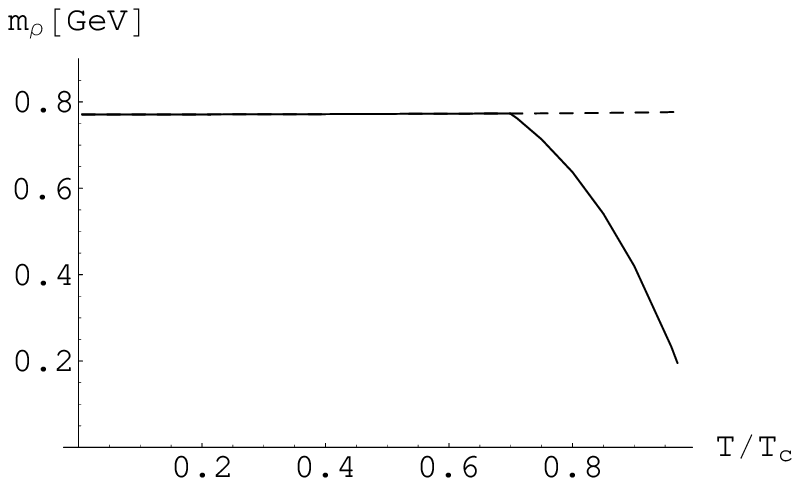}
\\
(a)
\\
\includegraphics[width = 7cm]{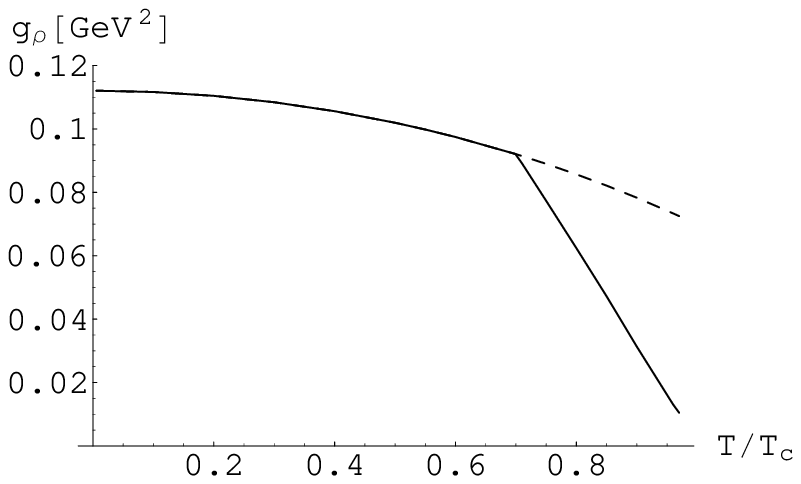}
\\
(b)
\end{center}
\caption{
Temperature dependences of (a) the vector meson mass $m_\rho$,
(b) the $\rho$-$\gamma$ mixing strength $g_\rho$.
The solid curves denote the full (both intrinsic
and hadronic) temperature dependences.
The curves with the dashed lines include only the hadronic
temperature effects. 
}
\label{fig:gr}
\end{figure}

We show the temperature dependences of the $\rho$ meson mass
$m_\rho(T)$ and the $\rho$-$\gamma$ mixing strength
at $\rho$ on-shell $g_\rho(s=m_\rho;T)$ in Fig.~\ref{fig:gr}.
In this figure
the dashed lines show the temperature dependences
of the physical quantities when only the hadronic corrections
are included, i.e., the parameters of the HLS Lagrangian are
assumed to have no temperature dependence.
While the solid lines show the full temperature dependences
by including the intrinsic temperature effects into the 
bare parameters as given in Eqs.~(\ref{Tdep g a}) and
(\ref{Tdep z F}).
Figure~\ref{fig:gr}(a) shows that
the vector meson mass including only the hadronic correction
little changes with temperature and the hadronic correction gives 
a positive contribution to $m_\rho$, 
$\delta^{\rm (had)} \simeq 5$ MeV.
In the temperature region above the flash temperature,
$T/T_c > T_f/T_c = 0.7$, the $\rho$ mass with the 
intrinsic effect rapidly drops
correspondingly to the rapid decreasing of the gauge coupling
as shown in Fig.~\ref{fig:bareT}.
In Fig.~\ref{fig:gr}(b) we can see that the hadronic effect
gives a negative correction 
to the $\rho$-$\gamma$ mixing strength at $\rho$ on-shell.
This is dominated by the decreasing of $f_\sigma$:
In the low-temperature region
Eq.~(\ref{fsig:T}) is approximated as
\begin{equation}
\left( f_\sigma^t(T) \right)^2
= f_\sigma^t(T)f_\sigma^s(T)
\simeq \bar{F}_\sigma^2 - \frac{N_f}{48} \, \bar{a}^2 T^2 \,,
\end{equation}
for $T \ll m_\rho$.
Above the flash temperature, the intrinsic effect causes the
rapid drop of the gauge coupling $g$, which further
decreases the $g_\rho$ toward zero.

\begin{figure}
\begin{center}
\includegraphics[width = 7cm]{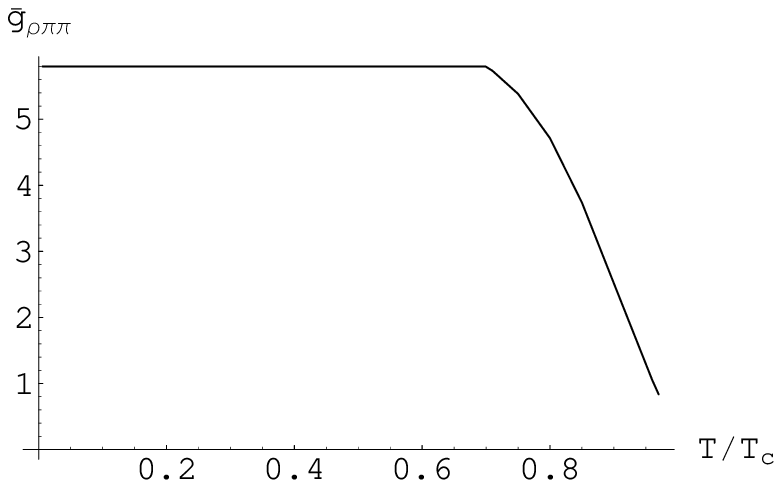}
\\
(a)
\\
\includegraphics[width = 7cm]{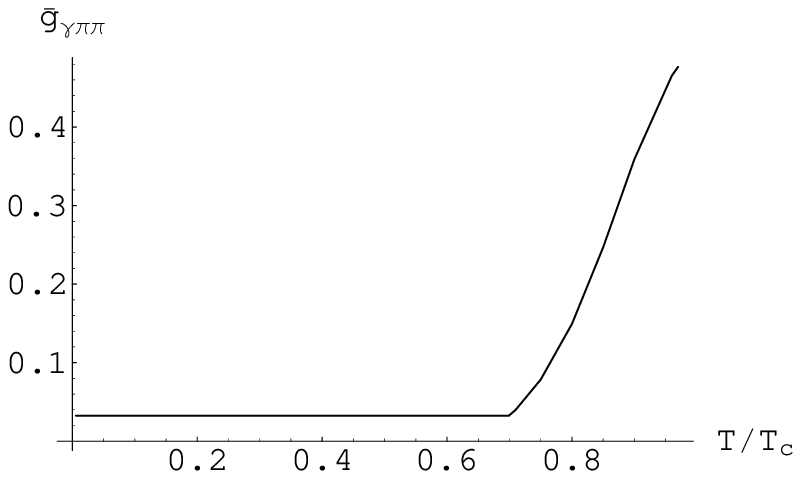}
\\
(b)
\end{center}
\caption[]{
Temperature dependences of 
(a) the parametric $\rho\pi\pi$ coupling
$\bar{g}_{\rho\pi\pi}$ and 
(b) the parametric direct-$\gamma\pi\pi$ coupling
$\bar{g}_{\gamma\pi\pi}$.
The lines denote the temperature dependences
including only the intrinsic effects.
}
\label{fig:gpp}
\end{figure}

As we show in the next section and Appendix~\ref{app:Imaginary},
the $\rho\pi\pi$ and the $\gamma\pi\pi$
couplings appearing in the imaginary part of the photon self-energy
do not have the hadronic temperature corrections in the present one-loop
calculation.
So we need the parametric $\rho\pi\pi$ coupling and the parametric
direct-$\gamma\pi\pi$ coupling.
By extending the form given in Eqs.~(\ref{grpp_vac})
and (\ref{gamma-pi-pi}),
they are given by
\begin{eqnarray}
&&
\bar{g}_{\rho\pi\pi}(T) = \bar{g}\,\frac{\bar{a}(0)}{2}\,,
\label{gpp}
\\
&&
\bar{g}_{\gamma\pi\pi}(T) = 1 - \frac{\bar{a}(0)}{2} \,.
\label{ggpp}
\end{eqnarray}
where $\bar{a}(0)$ is 
defined as~\cite{HS:VD}
\begin{equation}
\bar{a}(0) 
= \frac{\bar{F}_\sigma^2} {\bar{F}_\pi^2}
= \frac{F_\sigma^2(\mu=m_\rho;T)} {F_\pi^2(\mu=0;T)}\,.
\end{equation}
Here we should note that the barred quantities include only
the intrinsic effects.
In Fig.~\ref{fig:gpp},
we show that the temperature dependences of
$\bar{g}_{\rho\pi\pi}(T)$ and 
$\bar{g}_{\gamma\pi\pi}(T)$.
In Fig.~\ref{fig:gpp}(a) 
the rapid decrease of 
the parametric $\rho\pi\pi$ coupling $\bar{g}_{\rho\pi\pi}$
 occurs above the flash temperature
because of the drop of the gauge coupling $g$ caused by 
the intrinsic effect.
Figure~\ref{fig:gpp}(b) shows the temperature
dependence of the parametric direct-$\gamma\pi\pi$ coupling 
$\bar{g}_{\gamma\pi\pi}$.
In the temperature region below the flash temperature
$\bar{g}_{\gamma\pi\pi}$ is almost zero realizing the vector 
dominance (VD).
Above the flash temperature the parameter $\bar{a}(0)$ starts
to decrease from 2 to 1 due to the intrinsic effect
as given in Eq.~(\ref{Tdep g a}).
This causes an increase of $\bar{g}_{\gamma\pi\pi}$ toward $1/2$,
which implies the strong violation of the VD.


\setcounter{equation}{0}
\section{Form factor and dilepton spectra}
\label{sec:FF}

A lepton pair is emitted from the hot/dense matter
through a decaying virtual photon.
The differential production rate in the medium for fixed 
temperature $T$
is expressed in terms of the imaginary part of the photon 
self-energy $\mbox{Im}\Pi$ as
\begin{equation}
\frac{dN}{d^4q}(q_0,\vec{q};T)
=\frac{\alpha^2}{\pi^3 M^2}\frac{1}{e^{q_0/T}-1}
\mbox{Im}\Pi (q_0,\vec{q};T)\,,
\label{rate}
\end{equation}
where $\alpha = e^2/4\pi$ is the electromagnetic coupling constant,
$M$ is the invariant mass of the produced dilepton and 
$q_\mu=(q_0,\vec{q})$ denotes the momentum of the virtual photon.
The three-momentum integrated rate is given by
\begin{equation}
\frac{dN}{ds}(s;T) 
=\int\frac{d^3\vec{q}}{2q_0}\frac{dN}{d^4q}(q_0,\vec{q};T)\,,
\label{dl-rate}
\end{equation}
with $s = M^2$ and $q_0 = \sqrt{\vec{q}^2 + M^2}$.
We will focus on an energy region around the $\rho$ meson mass
scale in this analysis.
In this energy region it is natural to expect that
the photon self-energy is dominated by the two-pion process
and its imaginary part is related to the pion electromagnetic 
form factor ${\cal F}(s;T)$ through 
\begin{equation}
\mbox{Im}\Pi(s;T)
= \frac{1}{6\pi\sqrt{s}}
\left( \frac{s - 4m_\pi^2}{4} \right)^{3/2}
\left| {\cal F}(s;T) \right|^2\,,
\label{Im Pi}
\end{equation}
with the pion mass $m_\pi$.
By extending Eq.~(\ref{form factor-vac}) into the in-medium 
expression, the form factor is written as
\begin{eqnarray}
&&
{\cal F}(s;T)
= \bar{g}_{\gamma\pi\pi}(T)
\nonumber\\
&&\quad
{}+ \frac{g_\rho(s;T) \cdot \bar{g}_{\rho\pi\pi}(T)}
    {m_\rho^2(T) - s - \theta(s - 4m_\pi^2) 
     i m_\rho(T) \Gamma_\rho(s;T)}\,,
\nonumber\\
\label{form factor}
\end{eqnarray}
where we 
neglected the temperature
dependences of the pion mass.
The details are given in Appendix~\ref{app:Imaginary}.

\begin{figure}
\begin{center}
\includegraphics[width = 7cm]{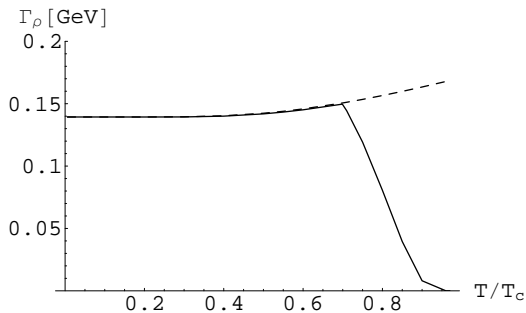}
\end{center}
\caption{
Decay width of the $\rho$ meson as a function of $T/T_c$
for $\sqrt{s}=m_\rho$.
The curve with the dashed line includes only the hadronic
temperature effects. The solid curve denotes the full (both intrinsic
and hadronic) temperature dependences.
}
\label{fig:width}
\end{figure}

Using the in-medium parameters obtained in the previous
section, we calculate the thermal width of the $\rho$ meson.
We present the temperature dependence of the width in
Fig.~\ref{fig:width}. 
The dashed line shows the temperature dependence of $\Gamma_\rho$
when only the hadronic effect is included.
Since $\bar{g}_{\rho\pi\pi}$ is independent of $T$ and
$m_\rho$ slightly increases with $T$ as shown in Fig.~\ref{fig:gr}(a)
(dashed line), $\Gamma_\rho$ increases with $T$ 
which implies that
the hadronic effect causes the broadening of the $\rho$ width.
The solid line in Fig.~\ref{fig:width} shows the temperature
dependence of $\Gamma_\rho$ when the intrinsic effect is also
included for $T>T_f$.
In this case $\bar{g}_{\rho\pi\pi}$ as well as $m_\rho$ decrease
with $T$ as $\bar{g}$ in the VM, and
the width $\Gamma_\rho$ decrease as
$\Gamma_\rho \sim \bar{g}^3 \rightarrow 0$.

\begin{figure}
\begin{center}
\includegraphics[width = 7cm]{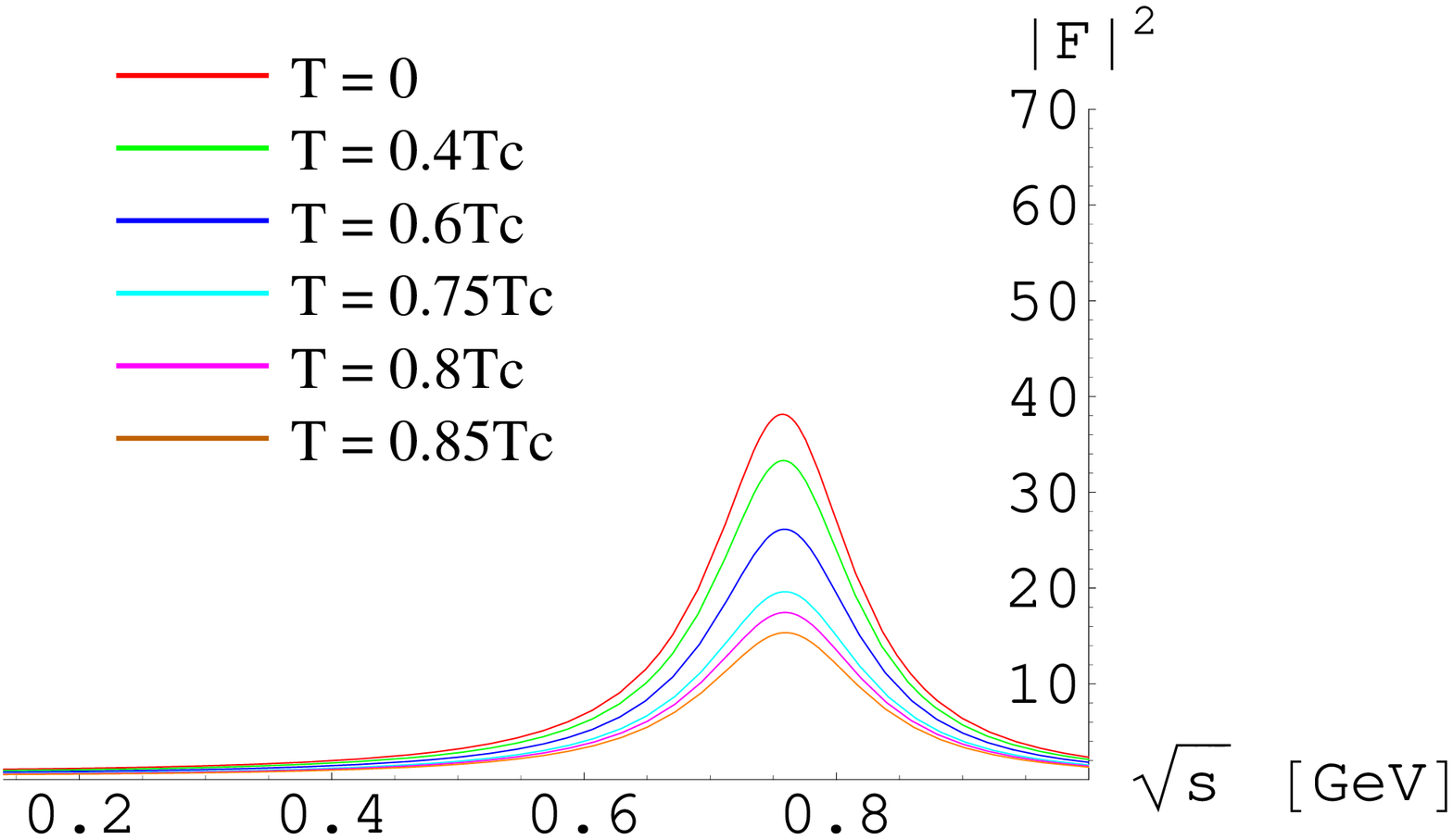}
\\
(a)
\\
\includegraphics[width = 7cm]{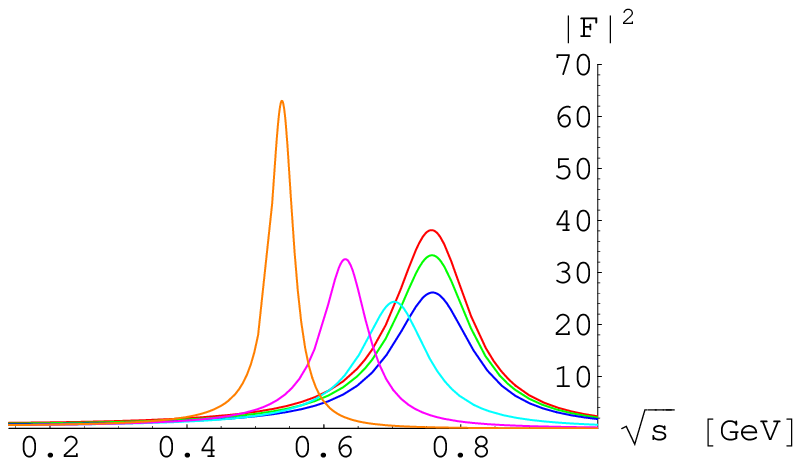}
\\
(b)
\end{center}
\caption{
Electromagnetic form factor of the pion as a function of
the invariant mass $\sqrt{s}$ for several temperatures.
The curves in the upper panel (a) include only the hadronic
temperature effects and those in the lower panel (b) include
both intrinsic and hadronic
temperature effects.
}
\label{fig:form}
\end{figure}

Now, let us calculate the pion form factor and 
the dilepton production rate.
Figure~\ref{fig:form} shows the electromagnetic 
form factor ${\cal F}$
given in Eq.~(\ref{form factor}) for several temperatures.
In Fig.~\ref{fig:form} (a) $g_\rho(s)$, $m_\rho$ and $\Gamma_\rho$ 
in the form factor include
only the hadronic temperature corrections and $\bar{g}_{\gamma\pi\pi}$
and $\bar{g}_{\rho\pi\pi}$ have no temperature dependence.
There is no remarkable shift of the $\rho$ meson mass
but the width becomes broader with increasing temperature, 
which is consistent with the previous study~\cite{SK}.
In Fig.~\ref{fig:form} (b) the intrinsic temperature effect are also
included into all the parameters in the form factor:
$g_\rho(s)$, $m_\rho$ and $\Gamma_\rho$ include the intrinsic
effect in addition to the hadronic one, and 
$\bar{g}_{\gamma\pi\pi}$ and $\bar{g}_{\rho\pi\pi}$
include the intrinsic one.
At the temperature below $T_f$, 
the hadronic effect dominates the form factor,
so that the curves for $T = 0$, $0.4T_c$ and $0.6T_c$
agree with the corresponding ones in Fig.~\ref{fig:form}(a).
At $T = T_f$ the intrinsic effect starts to contribute
and thus in the temperature region above $T_f$ 
the peak position of the form factor moves as
$m_\rho(T) \rightarrow 0$ with increasing temperature toward $T_c$.
Associated with this dropping $\rho$ mass,
the width becomes narrow, and 
the value of the form factor at the peak
grows up as~\cite{HS:VM}
\begin{eqnarray}
\left\vert \frac{g_\rho \bar{g}_{\rho\pi\pi}}{m_\rho\Gamma_\rho}
\right\vert^2 
\sim 
\left( \frac{g_\rho}{\bar{g}_{\rho\pi\pi}m_\rho^2} \right)^2
\sim 
\frac{1}{\bar{g}^2}
\ .
\end{eqnarray}

\begin{figure}
\begin{center}
\includegraphics[width = 7cm]{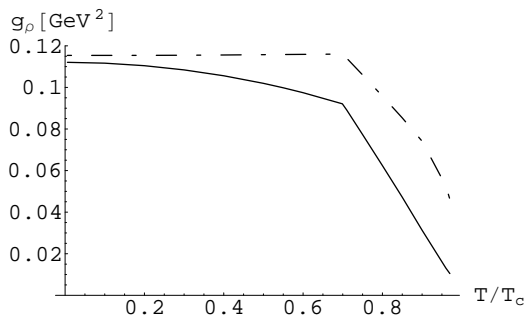}
\end{center}
\caption{
Temperature dependence of the $\rho$-$\gamma$ mixing strength 
$g_\rho$ for $\sqrt{s}=m_\rho$.
The dash-dotted curve corresponds to the case with the VD 
assumption.
The solid one includes the effect of the VD violation 
due to the VM.
}
\label{fig:gr-vd}
\end{figure}

As noted in subsection~\ref{ssec:VM},
the vector dominance (VD)
is controlled by the parameter $a$ in the HLS theory.
The VM leads to the strong violation of the VD 
(indicated by ``$\Slash{\rm VD}$'') 
near the chiral symmetry restoration point, which can be traced 
through the Wilsonian matching and the RG evolutions.
When the VD is assumed to be valid even in hot matter, i.e., 
$g_{\gamma\pi\pi}=0$,
one obtains the following constraint on $g_\rho$ for $s=0$
by imposing the normalization of the electromagnetic charge:
\begin{equation}
g_\rho^{\rm (VD)}(s;T)
= \frac{m_\rho^2(T)}{\bar{g}_{\rho\pi\pi}(T)} 
{}- s \, \bar{g}(T) \bar{z}_3(T)\,.
\end{equation}
Figure~\ref{fig:gr-vd} shows the temperature dependence of
the $\rho$-$\gamma$ mixing strength $g_\rho$ at $\rho$ on-shell
with VD (dash-dotted line) and $\Slash{\rm VD}$ (solid line).
In the low-temperature region, $T<T_f$,
the hadronic corrections to $m_\rho$ and $\bar{g}_{\rho\pi\pi}$ 
are small, so that the $\rho$-$\gamma$ mixing strength
with VD, $g_\rho^{\rm(VD)}$,
is almost stable against the temperature (see the dash-dotted line).
While $g_\rho$ with $\Slash{\rm VD}$ gets a non-negligible
hadronic correction in the HLS, which causes a decreasing
against the temperature (see the solid line).
Near the critical temperature, $T>T_f$, on the other hand,
both $m_\rho$ and $\bar{g}_{\rho\pi\pi}$ drop due to the VM and
the above ratio also decreases since $m_\rho^2/\bar{g}_{\rho\pi\pi}
\propto \bar{g}$.
However compared to the $g_\rho$ with $\Slash{\rm VD}$,
the decreasing of $g_\rho^{\rm(VD)}$ (dash-dotted line)
is much more gentle.
This affects the pion form factor which exhibits a strong 
suppression provided by decreasing $g_\rho$ in the VM.

\begin{figure*}
\begin{center}
\includegraphics[width = 7cm]{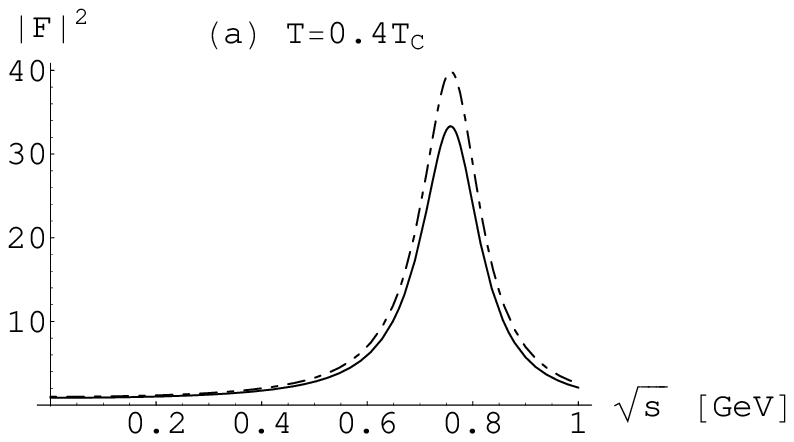}
\hspace*{1cm}
\includegraphics[width = 7cm]{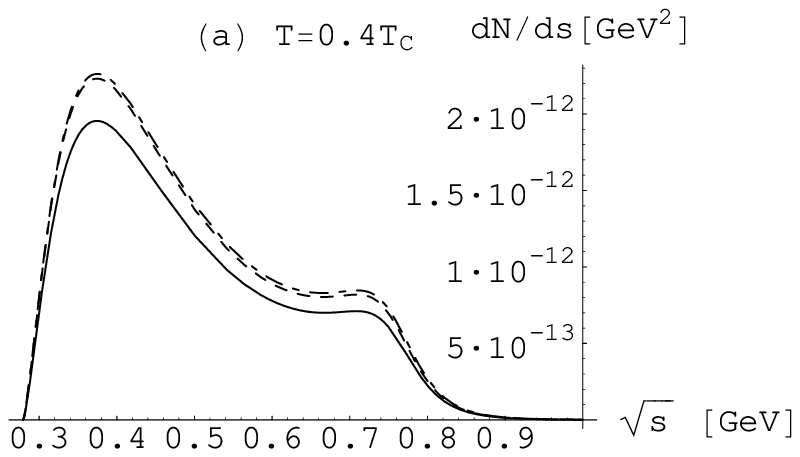}
\\
\includegraphics[width = 7cm]{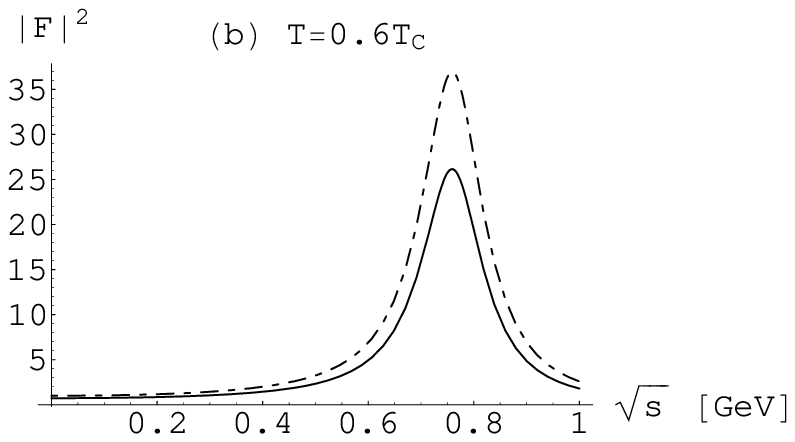}
\hspace*{1cm}
\includegraphics[width = 7cm]{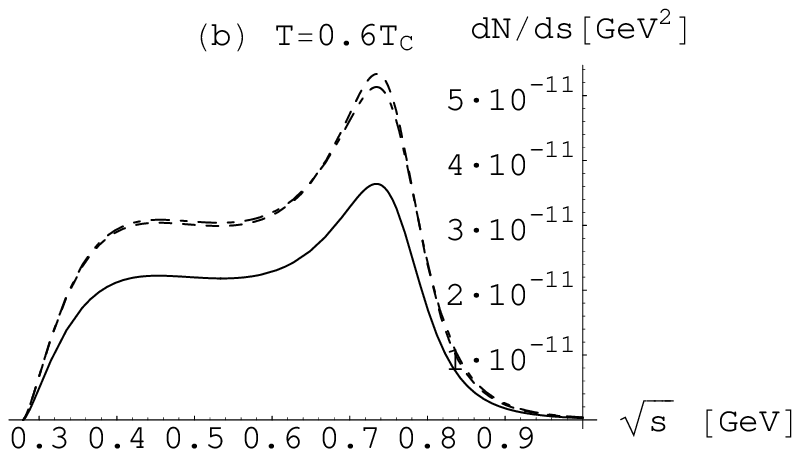}
\\
\includegraphics[width = 7cm]{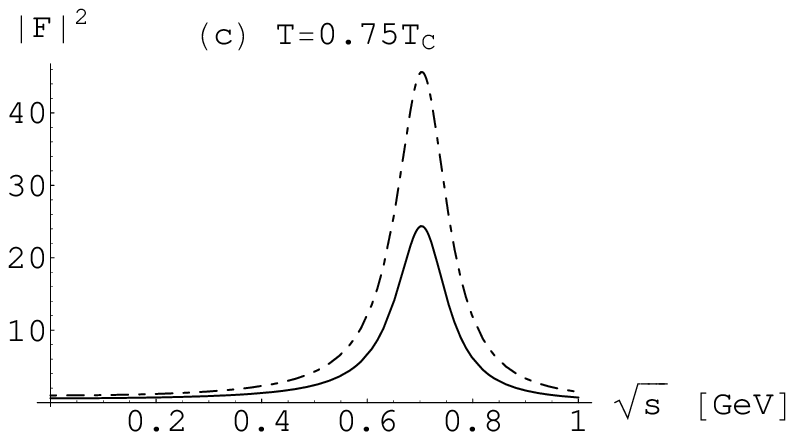}
\hspace*{1cm}
\includegraphics[width = 7cm]{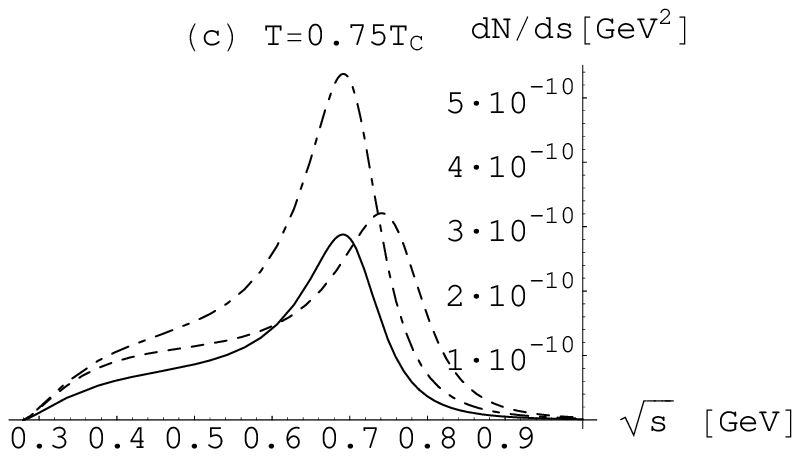}
\\
\includegraphics[width = 7cm]{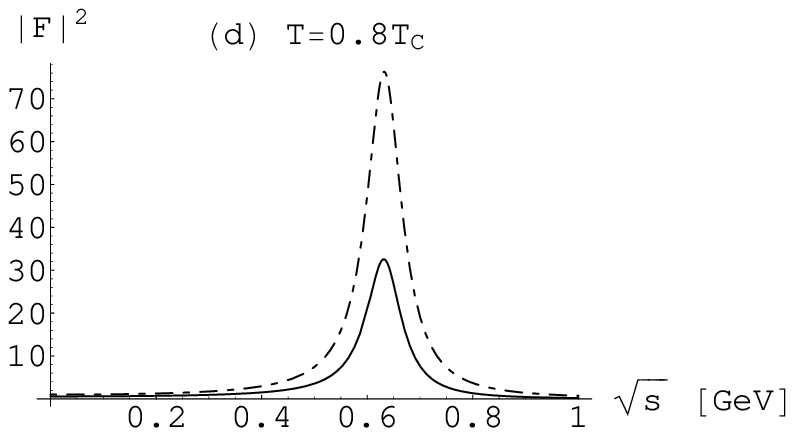}
\hspace*{1cm}
\includegraphics[width = 7cm]{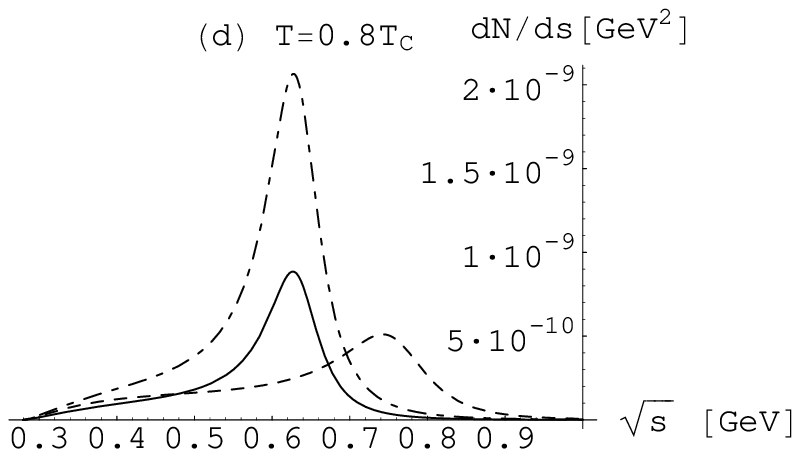}
\\
\includegraphics[width = 7cm]{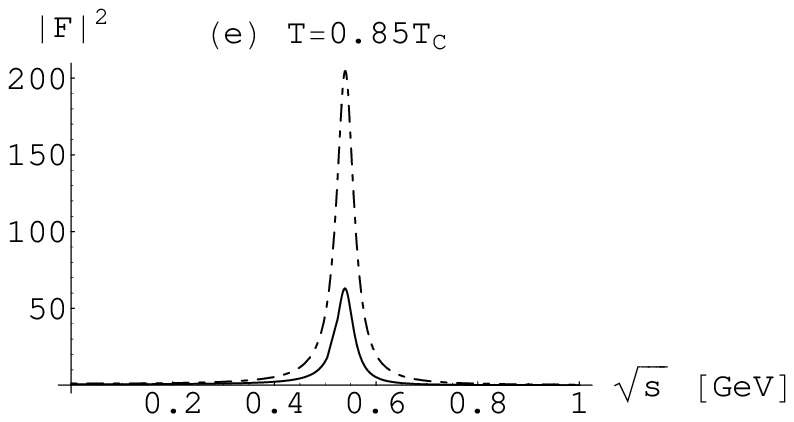}
\hspace*{1cm}
\includegraphics[width = 7cm]{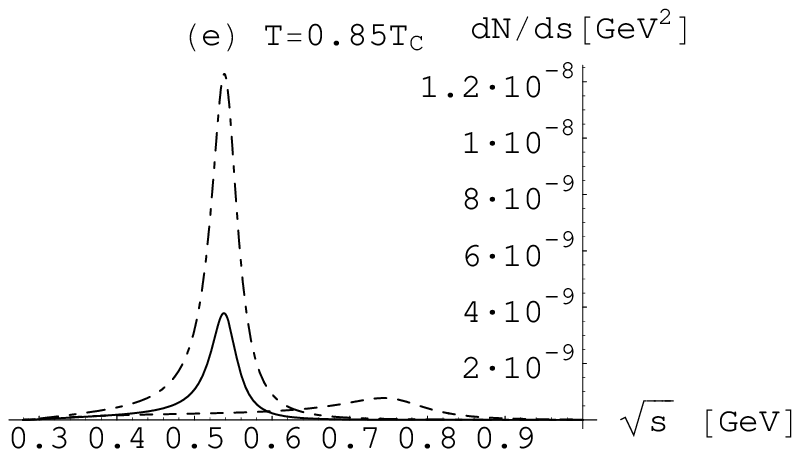}
\end{center}
\caption{
Electromagnetic form factor of the pion (left) and
dilepton production rate (right) as a function of the invariant 
mass $\sqrt{s}$ for various temperatures.
The solid lines include the effects of the violation of the VD.
The dashed-dotted lines correspond to the analysis assuming
the VD. In the dashed curves in the right-hand figures, 
the parameters at zero temperature were used.
}
\label{fig:dl}
\end{figure*}

Figure~\ref{fig:dl} shows the form factor and the dilepton 
production rate integrated over three-momentum,
in which the results with VD and $\Slash{\rm VD}$
were compared.
The figure shows a clear difference between the curves
with VD and $\Slash{\rm VD}$.
In the low-temperature region
$T \ll T_f$,
the hadronic effects are dominant compared with the intrinsic ones, 
so both curves almost coincide.
A difference between them starts to appear around $T=T_f$
and increases with 
temperature.
It can be easily seen that the $\Slash{\rm VD}$ gives a reduction
compared to the case with keeping the VD.
The features of the form factor as well as the dilepton production 
rate coming from two-pion annihilation shown in Fig~\ref{fig:dl}
(a)-(e) are summarized below for each temperature:
\begin{description}
\item[(a) and (b) (below $T_f$) :]
The form factor, which has a peak at the $\rho$ meson mass
$\sqrt{s} \sim 770$ MeV, is slightly suppressed with 
increasing temperature.
An extent of the suppression in case with $\Slash{\rm VD}$ is 
greater than that with VD.
This is due to decreasing of the $\rho$-$\gamma$ mixing strength
$g_\rho$ at finite temperature [see Fig.~\ref{fig:gr-vd}].
At $T < T_f$, $g_\rho$ mainly decreases by hadronic corrections.
In case with VD, however, $g_\rho^{\rm (VD)}$ is almost constant.
The dilepton rate (a) has two peaks, one is at the $\rho$ meson mass
and another one is lying around low-mass region.
The later peak comes from the Boltzman factor of Eq.~(\ref{rate}).
For a rather low-temperature the production rate is much enhanced
compared with the $\rho$ meson peak
since the yield in the higher-mass region is suppressed by 
the statistical factor.
With increasing temperature those peaks of the production rate (b)
are enhanced and the peak at $\sqrt{s} \sim m_\rho$ clearly appears.
In association with decreasing $g_\rho$, one sees a reduction
of the dilepton rate with $\Slash{\rm VD}$.
\item[(c), (d) and (e) (above $T_f$) :]
Since the intrinsic temperature effects are turned on, a shift of
the $\rho$ meson mass to lower-mass region can be seen.
Figure~\ref{fig:gr-vd} shows that $g_\rho$ is further reduced
by the intrinsic effects and much more rapidly decreases
than $g_\rho^{\rm (VD)}$.
Thus the form factor, which becomes narrower with increasing
temperature due to the dropping $m_\rho$, exhibits an obvious
discrepancy between the cases with VD and $\Slash{\rm VD}$.
The production rate based on the VM 
(i.e., the case with $\Slash{\rm VD}$) is suppressed compared
to that with the VD.
We observe that the suppression is more transparent 
for larger temperature:
The suppression factor is $\sim 1.8$ in (c), $\sim 2$ in (d)
and $\sim 3.3$ in (e). 

As one can see in (c), the peak value of the rate
predicted by the VM
in the temperature region slightly above the flash temperature
is even smaller than the one obtained by the vacuum parameters,
and the shapes of them are quite similar to each other.
This indicates that it might be difficult to measure the 
signal of the dropping $\rho$ experimentally, if this
temperature region is dominant in the evolution of the fireball.
In the case shown in (d), on the other hand,
the rate by VM 
is enhanced by a factor of 
about two compared with the one by the vacuum $\rho$.
The enhancement becomes prominent near the critical temperature
as seen in (e).
These imply that we may have a chance to discriminate the
dropping $\rho$ from the vacuum $\rho$.

\end{description}


\setcounter{equation}{0}
\section{Summary and Discussions}
\label{sec:sum}

We studied the pion electromagnetic form factor and the thermal
dilepton production rate from the two-pion annihilation
within the hidden local symmetry (HLS) theory as an effective
field theory of low-energy QCD.
In the HLS theory the chiral symmetry is restored as the vector
manifestation (VM) in which the massless $\rho$ meson joins
the same chiral multiplet as pions.
In order to determine the temperature dependences of the
parameters of the HLS Lagrangian, the Wilsonian matching to
the operator product expansion at finite temperature was made
by applying the matching scheme developed 
in the vacuum~\cite{HY:WM,HY:PRep} and at the critical 
temperature~\cite{HS:VM,HKRS:SUS,HS:VD,PiV}.

In the notion of the Wilsonian matching to define a bare theory
in hot environment,
the bare parameters are dependent on temperature, which
are referred as the intrinsic temperature effects.
At low temperatures the chiral properties of in-medium hadrons
are dominated by ordinary hadronic loop corrections.
The dropping $\rho$ is realized in the HLS framework due to
the intrinsic effects and thus they play crucial roles especially
near the chiral phase transition.
In order to see an influence of the intrinsic temperature effects,
we presented the form factor including full temperature effects,
i.e., the intrinsic and hadronic effects, and compared
with that including only hadronic corrections.
The $\rho$ meson mass $m_\rho$ is almost stable against
the hadronic corrections and one does not obtain the dropping $m_\rho$.
Accordingly the peak of the form factor including only the 
hadronic effects is located at around $\sqrt{s} \sim m_\rho \sim 770$
MeV even at finite temperature.
The form factor is reduced with increasing temperature and
correspondingly becomes broader.
On the other hand, the Wilsonian matching procedure certainly 
involves the intrinsic temperature effects in the analysis
and provides the dropping $m_\rho$ as the VM.
The form factor {\it above the flash temperature $T_f$} thus 
starts to present a shift of $m_\rho$ to lower invariant mass region.
Associated with the dropping $\rho$, the form factor becomes sharp.

One of the significant predictions of the VM is a strong violation 
of the vector dominance (VD) of the pion form factor.
The VM predicts that the VD is maximally violated at the transition
temperature and it crucially affects the analysis of dilepton yields.
We presented the form factor and the dilepton production rate
with and without the VD assumption together with the dropping $\rho$.
For $T \ll T_f$ the result shows only 
a small difference between those
two cases since the VD is still well satisfied in low temperatures.
A clear difference can be seen for $T > T_f$ where the intrinsic
temperature effects contribute to the physical quantities.
The form factor and consequently the dilepton production rate
with taking account of the violated VD are reduced and 
exhibit an obvious difference near $T_c$ compared to those 
with the VD.
Such behaviors can be understood from the in-medium $\rho$-$\gamma$
mixing strength $g_\rho$.
At $T \ll T_f$, $g_\rho$ mainly decreases due to the hadronic 
corrections and does not show a big difference.
With increasing temperature but still below $T_f$,
$g_\rho$ shows a discrepancy between the cases with/without the VD,
which is roughly 17\% around $T = 0.6\,T_c$.
The intrinsic temperature effect comes in at $T_f$ and provides a more
rapid decreasing of $g_\rho$ than $g_\rho^{\rm (VD)}$.
This variation reaches about 40\% at $T = 0.85\,T_c$ and 
eventually causes a precise difference of the production rate 
by factor $\sim 3.3$.

Several comments are in order:

The HLS Lagrangian has only pions and vector mesons as physical
degrees of freedom, and a time evolution was not considered
in this work.
Thus it is not possible to make a direct comparison of our results
with experimental data.
However a {\it naive} dropping $m_\rho$ formula, i.e., $T_f = 0$, 
as well as VD in hot/dense matter are sometimes used for
theoretical implications of the data.
As we have shown in this paper, the intrinsic temperature effects
together with the violation of the VD give a clear difference
from the results without including those effects.
It may be then expected that a field theoretical analysis
of the dropping $\rho$ as presented in this work and a reliable
comparison with dilepton measurements will provide an evidence
for the in-medium hadronic properties associated with the chiral
symmetry restoration, if complicated hadronization processes do
not wash out those changes.

Our analysis can be applied to a study at finite density.
Especially to study under the conditions for SPS and
future GSI/FAIR should be addressed as an important issue.
In such a dense environment, the particle-hole configurations
with same quantum numbers with pions and $\rho$ mesons are 
crucial~\cite{FP}.
The violation of the VD has been also presented at finite density
in the HLS theory~\cite{HKR:VM}.
Therefore the dilepton rate as well as the form factor will be
much affected by the intrinsic density effects and be reduced
above the ``flash density''.

Recently the chiral perturbation theory with including vector
and axial-vector mesons as well as pions has been constructed%
~\cite{HS:GHLS,hidaka} based on the generalized HLS%
~\cite{BKY:NPB,BKY:PRep,BFY:GHLS}.
In this theory the dropping $\rho$ and $A_1$ meson masses
were formulated and it was shown that the dropping masses
are related to the fixed points of the RGEs which gives
a VM-type restoration and that the VD is strongly violated
also in this case.
Inclusion of the effect of $A_1$ meson as well as
the effect of collisional broadening will be done in future
work.


\subsection*{Acknowledgments}

We are grateful to Gerry Brown, Bengt Friman and Mannque Rho 
for fruitful discussions and comments.
We also thank Jochen Wambach for stimulating discussions.
The work of C.S. was supported in part by the Virtual Institute
of the Helmholtz Association under the grant No. VH-VI-041.
The work of M.H. 
is supported in part by the Daiko Foundation \#9099, 
the 21st Century
COE Program of Nagoya University provided by Japan Society for the
Promotion of Science (15COEG01), and the JSPS Grant-in-Aid for
Scientific Research (c) (2) 16540241.


\appendix

\setcounter{section}{0}
\renewcommand{\thesection}{\Alph{section}}
\setcounter{equation}{0}
\renewcommand{\theequation}{\Alph{section}.\arabic{equation}}

\section{Functions}
\label{app:functions}

In this appendix, we list the integral forms of the functions 
which appear in the expressions of the hadronic corrections.
The functions $I_{n}(T), J^n_{m}(M;T), F^n_{3}(p_0;M;T)$ and 
$G_{n}(p_0;T)$ ($n$, $m$: integers) are given by
\begin{eqnarray}
&&
I_{n}(T) 
= \int \frac{d^3 k}{(2\pi)^3}\frac{|\vec{k}|^{n-3}}{e^{k/T}-1}\,, 
\nonumber\\
&&
J^n_{m}(M;T) 
= \int \frac{d^3 k}{(2\pi)^3} \frac{1}{e^{\omega /T}-1}
\frac{|\vec{k}|^{n-2}}{{\omega}^m}\,, 
\nonumber\\
&&
F^n_{3}(p_0;M;T) 
= \int \frac{d^3 k}{(2\pi)^3}\frac{1}{e^{\omega /T}-1}
\frac{4|\vec{k}|^{n-2}}{\omega (4{\omega}^2 - {p_0}^2)}\,, 
\nonumber\\
&&
G_{n}(p_0;T) 
= \int \frac{d^3 k}{(2\pi)^3}\frac{|\vec{k}|^{n-3}}{e^{k/T}-1}
\frac{4|\vec{k}|^2}{4|\vec{k}|^2 - {p_0}^2}\,,
\end{eqnarray}
where we define
\begin{equation}
\omega = \sqrt{k^2 + {M}^2}\,.
\end{equation}


\section{Imaginary Part of the Photon Self-Energy}
\label{app:Imaginary}

In this appendix, we show the imaginary part of the photon
self-energy obtained in the HLS.
The photon self-energy used in section~\ref{sec:FF} is
related to the vector current correlator in the HLS as
\begin{equation}
\mbox{Im}\,\Pi(s) = s \, \mbox{Im}\,G_V(s) \,.
\end{equation}
Let $\Pi_{V}^{\mu\nu}$, $\Pi_{V\parallel}^{\mu\nu}$ and
$\Pi_{\parallel}^{\mu\nu}$ denote
the $V$-$V$, $V$-${\mathcal V}$ and 
${\mathcal V}$-${\mathcal V}$ two point functions, respectively.
These two-point functions are decomposed 
as~\cite{HY:PRep,Sasaki:D}
\begin{equation}
\Pi^{\mu\nu}(q) = g^{\mu\nu} \Pi^S(q^2) + 
  \left( g^{\mu\nu} q^2 - q^\mu q^\nu \right) \Pi^{LT}(q^2)
\,.
\end{equation}
By using the two-point functions,
the vector current correlator $G_V(s)$ is expressed 
as~\cite{HY:PRep,Sasaki:D}
\begin{equation}
G_V = 
\frac{\Pi_V^S \left( - \Pi_V^{LT} - 2 \Pi_{V\parallel}^{LT} \right) }
 { \Pi_V^S + s \Pi_V^{LT} }
- \Pi_{\parallel}^{LT}
\,.
\end{equation}
By noting that $\Pi_V^S$ does not have imaginary part for 
$s < 4 m_\rho^2$ at one-loop, the imaginary part of the 
current correlator is expressed as
\begin{eqnarray}
&& \mbox{Im}\, G_V =
- \mbox{Im}\, \Pi_V^{LT} \,
\frac{\left(\Pi_V^S - s\,\mbox{Re}\,\Pi_{V\parallel}^{LT}\right)^2}
  { \left\vert \Pi_V^S + s\,\Pi_{V}^{LT} \right\vert^2 }
\nonumber\\
&& \quad {}- \mbox{Im}\,\Pi_{V\parallel}^{LT} \,
  \left[
    \frac{\Pi_V^S - s\,\mbox{Re}\,\Pi_{V\parallel}^{LT} }
      {\Pi_V^S + s\,\Pi_{V}^{LT} }
    +
    \frac{\Pi_V^S - s\,\mbox{Re}\,\Pi_{V\parallel}^{LT} }
      {\left(\Pi_V^S + s\,\Pi_{V}^{LT}\right)^{\ast} }
  \right]
\nonumber\\
&& \quad {} - \mbox{Im}\,\Pi_{\parallel}^{LT}
\,,
\label{Im G}
\end{eqnarray}
up to higher order terms such as
$\left( \mbox{Re}\,\Pi_{V\parallel}^{LT} \right)^2$
and $\left( \mbox{Re}\,\Pi_{V\parallel}^{LT} \right)
\left( \mbox{Im}\,\Pi_{V\parallel}^{LT} \right)$.

Using the formulas shown in Refs.~\cite{HY:PRep,Sasaki:D},
we obtain
\begin{eqnarray}
\mbox{Im}\,\Pi_V^{LT}(s) &=&
  - \frac{1}{48\pi} \left( \frac{s-4m_\pi^2}{s} \right)^{3/2}
  \, \left( \frac{\bar{g}_{\rho\pi\pi}}{\bar{g}} \right)^2
\,,
\nonumber\\
\mbox{Im}\,\Pi_{V\parallel}^{LT}(s) &=&
  - \frac{1}{48\pi} \left( \frac{s-4m_\pi^2}{s} \right)^{3/2}
  \, \bar{g}_{\gamma\pi\pi}\,\frac{\bar{g}_{\rho\pi\pi}}{\bar{g}}
\,,
\nonumber\\
\mbox{Im}\,\Pi_{\parallel}^{LT}(s) &=&
  - \frac{1}{48\pi} \left( \frac{s-4m_\pi^2}{s} \right)^{3/2}
  \left( \bar{g}_{\gamma\pi\pi} \right)^2
\,.
\end{eqnarray}
Here $\bar{g}_{\rho\pi\pi}$ and $\bar{g}_{\gamma\pi\pi}$
are expressed by the parameters of the Lagrangian as in 
Eqs.~(\ref{grpp_vac}) and (\ref{gamma-pi-pi}).
We put the bar to clarify that the barred quantities
do not include the hadronic temperature corrections
and include only the intrinsic effects, when they are
considered in hot matter.
We define the $\rho$ propagator $D_\rho(s)$ and the
momentum-dependent $\rho$-$\gamma$ mixing strength $g_\rho(s)$ as
\begin{eqnarray}
D_\rho(s) &\equiv& \frac{1}
  {\bar{g}^2\, \left( \Pi_V^S(s) + s\,\Pi_{V}^{LT}(s) \right) }
\,
\nonumber\\
g_\rho(s) &\equiv& \bar{g} 
 \left(\Pi_V^S(s) - s\,\mbox{Re}\,\Pi_{V\parallel}^{LT}(s)\right)
\,.
\end{eqnarray}
In the present analysis, we consider the energy region around 
the $\rho$ meson mass, $s \sim m_\rho^2$.
Then we use the following approximate form for the $\rho$ propagator:
\begin{equation}
D_\rho(s) \simeq \frac{1}
  {m_\rho^2 -s - \theta(s-4m_\pi^2)\, i m_\rho \Gamma_\rho}
\,,
\label{app:rho prop}
\end{equation}
where both $m_\rho$ and $\Gamma_\rho$ include the hadronic
temperature effects in addition to intrinsic ones.
We should note that
there is not substantial momentum 
dependence in $\Pi_V^S(s)$ for $s < 4 m_\rho^2$,
since the pion loop does not contribute to $\Pi_V^S(s)$ except
for the tad-pole diagram.
Then, in the $\rho$-$\gamma$ mixing strength $g_\rho(s)$, 
we take $\Pi_V(s) \rightarrow \Pi_V(s=0) = f_\sigma^2$.
Furthermore, we neglect the momentum dependence of 
$\mbox{Re}\,\Pi_{V\parallel}^{LT}(s)$ for the consistency with the
approximation adopted in Eq.~(\ref{app:rho prop}).
This implies 
\begin{equation}
g_\rho(s) \simeq \bar{g}\, 
  \left( f_\sigma^2 - s\, \bar{z}_3 \right)
\,,
\label{app:grho}
\end{equation}
where $\bar{z}_3$ includes the intrinsic temperature effect.

By using the above quantities, 
the imaginary part of the current correlator $\mbox{Im}\,G_V$ 
in Eq.~(\ref{Im G}) is expressed as
\begin{eqnarray}
\mbox{Im}\,G_V(s) &=&
\frac{1}{48\pi} \left( \frac{s-4m_\pi^2}{s} \right)^{3/2}
\nonumber\\
&& \times \left\vert
  \bar{g}_{\gamma\pi\pi} + 
  g_\rho(s)\,\bar{g}_{\rho\pi\pi} \, D_\rho(s)
\right\vert^2
\,,
\end{eqnarray}
which leads to Eq.~(\ref{Im Pi}).


\end{document}